\documentclass[%
reprint,
 amsmath,amssymb,
 aps,
]{revtex4-2}

\usepackage{caption}
\usepackage{graphicx}
\usepackage{dcolumn}
\usepackage{bm}

\usepackage{amsmath}	
\usepackage{orcidlink}
\usepackage[margin=0.3cm]{caption}
\usepackage{subcaption}
\usepackage{hyperref}
\graphicspath{ {./Figs/} }
\usepackage{booktabs}
\usepackage{multirow}
\usepackage{float}
\restylefloat*{figure}
\restylefloat*{table}
\usepackage{algorithm} 
\usepackage{algpseudocode} 
\usepackage{enumitem}
\usepackage[utf8]{inputenc}
\usepackage[algo2e,ruled,vlined,linesnumbered,resetcount]{algorithm2e}

\usepackage{soul}

\usepackage{apssamp}
\usepackage{natbib, url}

\begin{document}

\preprint{APS/123-QED}

\title[Autoregressive Search of Gravitational Wave]{Autoregressive Search of Gravitational Waves: Denoising}

\author{Sangin Kim$^{1}$}
    \thanks{E-mail: kimsanginn@gmail.com}
\author{C. Y. Hui$^{1}$}
    \thanks{E-mail: cyhui@cnu.ac.kr, huichungyue@gmail.com}
\author{Jianqi Yan$^{2}$}  
\author{Alex P. Leung$^{3}$}    
\author{Kwangmin Oh$^{4}$} 
\author{A. K. H. Kong$^{5}$} 
\author{L.C.-C.~Lin$^{6}$}  
\author{Kwan-Lok Li$^{6}$}  

\affiliation{$^{1}$Department of Astronomy and Space Science, Chungnam National University, 9 Daehak-ro, Yuseong-gu, Daejeon 34134, Republic of Korea}

\affiliation{$^{2}$Faculty of Innovation Engineering, Macau University of Science and Technology, Avenida Wai Long, Taipa 999078, Macau}
\affiliation{$^{3}$Department of Physics, The University of Hong Kong, 999077, Hong Kong}
\affiliation{$^{4}$Department of Physics and Astronomy, Michigan State University, East Lansing, MI 48824, USA}
\affiliation{$^{5}$Institute of Astronomy, National Tsing Hua University, Hsinchu 30013, Taiwan}
\affiliation{$^{6}$Department of Physics, National Cheng Kung University, No.1, University Road, Tainan City 701, Taiwan}


\date{\today}

\begin{abstract}
Because of the small strain amplitudes of gravitational-wave (GW) signals, unveiling them in the presence of detector/environmental noise is challenging. For visualizing the signals and extracting its waveform for a comparison with theoretical prediction, a frequency-domain whitening process is commonly adopted for filtering the data. In this work, we propose an alternative template-free framework based on autoregressive modeling for denoising the GW data and extracting the waveform. 
We have tested our framework on extracting the injected signals from the simulated data as well as a series of known compact binary coalescence (CBC) events from the LIGO data. Comparing with the conventional whitening procedure, our methodology generally yields improved cross-correlation and reduced root mean square errors with respect to the signal model. 


\end{abstract}


\maketitle


\section{Introduction}
The existence of gravitational wave (GW) is one of the most remarkable predictions of general relativity (GR)\citep{1916SPAW.......688E,1918SPAW.......154E}. GW is a tidal acceleration that propagates in spacetime at the speed of light. According to the Einstein field equation, it requires stress at an order of $c^{2}/8\pi G\sim10^{43}$~N~m$^{-2}$ to produce a unit of curvature. Therefore, the amplitude of GW is expected to be very small. And it requires catastrophic phenomena involving compact objects to produce such tiny ripples in the spacetime (e.g. binary black hole mergers).

It is the small amplitude of GW that makes the detection challenging. The first compelling evidence for the existence of GW came indirectly from the long-term pulsar timing of the Hulse-Taylor binary PSR~B1913+16 \citep[][]{1975ApJ...195L..51H}. The behavior of this binary (e.g. decay of orbital period) is fully consistent with the prediction by GR as the system loses its orbital energy in GW.  

Thanks to the improved sensitivity, on 14 September 2015, the advanced Laser Interferometer Gravitational-wave Observatory (LIGO) has directly detected a GW event, GW150914, from a binary black hole (BBH) coalescence for the first time  \citep{PhysRevLett.116.061102}. This has opened the possibility of exploring our Universe without limiting to the window of electromagnetic radiation. Two years later, the era of multi-messenger astronomy was highlighted by the discovery of GW event GW170817 resulted from the merger of two neutron stars (NSs) \citep{PhysRevLett.119.161101,2017ApJ...848L..12A,2017ApJ...848L..13A}, which was found to be associated with the $\gamma-$ray burst GRB~170817A \citep{Goldstein_2017}. This marks the first case that both GW and electromagnetic radiation were detected from the same astrophysical object.

Currently, in the Gravitational Wave Transient Catalog (GWTC) maintained by LIGO/Virgo/KAGRA collaboration
\footnote{\href{https://gwosc.org/eventapi/html/allevents/}{https://gwosc.org/eventapi/html/allevents/}} 
 \citep[][]{2019PhRvX...9c1040A,2021PhRvX..11b1053A,2021arXiv210801045T,2021arXiv211103606T}, 
there are 93 GW transient events so far have been confidently detected (i.e. probability of origin from
an astrophysical source $p_{\rm astro}>0.5$). These include 89 from BBH coalescence, 2 from NS-NS mergers, and 2 from BH-NS mergers. Apart from these confident events, there are $>20$ marginal candidates.


For further advancing GW astronomy, while enhancing the instrumental sensitivity is vitally important \citep[e.g.][]{2013NaPho...7..613A}, progress can also be achieved by improving the methodology of data processing and analysis. Currently, the standard search method for CBC events in the GW community is matched filtering \citep[cf.][]{Abbott_2020b}, which is done by cross-correlating a template of known waveform and the interferometer output at different time delays to produce a filtered output. With the signal-to-noise ratio (SNR) as the ratio of the value of the filtered output to the corresponding value root mean square value for the noise, it can be proved that a matched filter comprises the ratio of the template of the actual waveform to the spectral noise density of the interferometer can optimize the SNR under several assumptions \citep{2017PhRvD..95d2001M}. 

Although the technique of matched filtering has unveiled a considerable population of GW events as aforementioned, it has a number of limitations. For the matched filter to have optimal performance, the data have to fulfill the assumptions of wide-sense stationarity (WSS) and zero-means, which are generally not satisfied in the raw interferometric data. And most importantly, the construction of matched filter requires knowledge of waveform for the expected signal. However, the forms of GW signal from many possible sources are poorly modeled (e.g. highly eccentric BH binaries) or even unknown (e.g. fast radio bursts). In such cases, matched filtering technique cannot be employed. Even for the cases that the waveform can be determined such as circular BH binaries, this technique still requires the construction of a large template bank to cover a sufficiently large parameter space. A search over this extensive template bank by brute-force is computationally expensive. 

Furthermore, even though the technique of matched filtering is capable to detect the GW signals from CBC events with known waveform, it does not enable one to visualize the signal directly. For visualizing the GW signal from the data and extracting its waveform for a comparison with the prediction by numerical relativity \citep[e.g. Fig. 1 in][]{PhysRevLett.116.061102}, one must filter the raw time series with a bandpass filter for removing the data out of the detectors' most sensitive frequency band as well as apply the frequency-domain whitening process for suppressing the colored noises at low frequencies and the spectral lines resulted from instrumental/background effects. Frequency-domain whitening is a procedure to equalize the spectrum through dividing the Fourier coefficients by the estimate of the amplitude spectral density of the noise. While this is considered as a standard procedure and has been adopted for noise suppression in many works \citep[e.g.][]{Hu_2022,2021NatSR..1120507A}, it is still important to explore alternative techniques for de-noising and compare their performance with that of conventional whitening filter. For example, Tsukada et al. have proposed a time-domain whitening filter for optimizing latency in the CBC data analysis pipeline \citep{2018PhRvD..97j3009T}. 

In this paper, we explore the feasibility of a template-free method based on autoregressive modeling in filtering GW data. In Section II, we provide an overview of the methodology. In Section III, we will demonstrate the feasibility of our framework by a series of experiments. And we will summarize our results and provide an outlook for further development in Section IV.

\section{Methodology}
\label{sec:Methodology}
\subsection{Autoregression}
Time series data that we acquire in nature can be affected by various random processes and exhibit stochastic behaviors. 
Due to the high sensitivity of GW detectors, the raw data are typically corrupted with the various kinds of noise \citep[e.g.][]{Bahaadini_2018}, in which the assumptions for the matched filter to attain the optimal performance such as stationarity are generally not fulfilled. 
In our proposed framework,  we adopted an autoregressive (AR) approach in developing an efficient time-domain noise filtering scheme without any a priori knowledge on the noise. 

In a recent astronomical application of AR modeling, Caceres et al. have developed a methodology of the autoregressive planet search (ARPS) for treating a wide variety of stochastic processes so as to improve the search of transit signals by exoplanets in the residuals after noise reduction \cite{2019AJ....158...57C}. In exoplanet search, people are looking for small dips in the light curve resulting from a transit submerged by the much larger brightness variability of the parent stars. And the aperiodic colored noise in the photometric data is notoriously difficult to treat \citep{2017AJ....153....3C}. With a procedure based on AR, Caceres et al. have demonstrated the stellar variability can be identified and removed.  

We notice that the aforementioned challenge is shared by the GW astronomy, namely searching for the small strain amplitude of GW signals in the presence of instrumental/environmental noise with amplitude orders of magnitude larger. This comparison has motivated us to explore whether the AR technique can also be applied in extracting GW signals.

AR modeling can be applied to any dynamical system whose status in the present time has a dependence on its past status (i.e. autocorrelated behavior). The simplest model AR($p$) can be built by regression with the estimate at time $t$, $\hat{x}_t$, being modeled by the linear combination of past values $x_{t-i}$ plus a random noise term:
\begin{equation}
    \hat{x}_t = \sum^p_{i=1} a_i x_{t-i} + \epsilon_t
	\label{eq:AR}
\end{equation}
where $p$ is the order of AR model (i.e. the number of lags in the model), $a_i$ are the model parameters, $x_{t-i}$ is the $i$-th past data, and $\epsilon_t$ is the noise term distributed as a Gaussian with zero mean and unknown variance.

In the application of ARPS, the best-fit AR model can be treated as a good estimator of stochastic noise. By subtracting the model from the raw data, the residuals can be obtained as follow:
\begin{equation}
  x_{r} = x-\hat{x}
  \label{eq:resid}
\end{equation}
where $\hat{x}$ is the best-fit AR model on data $x$. From $x_{r}$, we can investigate whether there is any  astrophysically-interesting signals can be extracted \citep[see][]{2019AJ....158...57C,2019AJ....158...58C}. 


\subsection{Autoregressive integrated moving-average model (ARIMA)}
\label{sec:ARIMA} 
However, for an AR model to provide a legitimate description of the data, the time series is assumed to be stationary. For the time series with systematic trends, such data cannot be treated by the AR model. For converting a non-stationary time series into a stationary one, the differencing operation is found to be efficient (e.g. $x_{t}^{'}=x_{t}-x_{t-1}$ where $x_{t}^{'}$ is the differenced series obtained from the change between consecutive values in the original time series). With the backshift operator $B$ defined as $Bx_{t}=x_{t-1}$, the aforementioned process can be described as $(1-B)x_{t}$. This is known as first-order differencing. To generalize the process to a higher orders, the operation can be modified as
\begin{equation}
x_{d}=(1-B)^{d}x_{t} 
\label{eq:integrate}
\end{equation}

\noindent which is commonly referred to as {\it Integrated process} (I) of 
$d^{\rm th}$ order. The output $x_{d}$ can be modeled as a stationary time series. 

While AR model uses past values in a time series to predict the current value, a Moving Average model of order $q$, MA($q$), predicts the current value by a linear combination of past error terms
\begin{equation}
    \hat{x}_t = \sum^q_{j=1} b_i \epsilon_{t-j} + \epsilon_t
\label{eq:MA}
\end{equation}

\noindent where $\epsilon_{t-j}$ is the error term for the $j-$th time step in the time series $x_{t}$ and $b_{i}$ are the model parameter.  


Combining \autoref{eq:AR}, \autoref{eq:integrate} and, \autoref{eq:MA}, an AutoRegressive Integrated Moving-Average model (ARIMA) can be constructed as:
\begin{equation}
    (1-B)^{d}x_t = \sum^p_{i=1} a_i x_{t-i} + \sum^q_{j=1} b_i \epsilon_{t-j} + \epsilon_t
	\label{eq:ARIMA}
\end{equation}

\noindent with $a_{i}$ and $b_{i}$ estimated simultaneously for the whole time series by any optimization method (e.g. maximum likelihood). On the other hand, the orders of the model (e.g. $p$, $q$, $d$) can be determined by the procedures of model selection. 

By applying ARIMA modeling to the light curves of 156717 stars as observed by NASA's {\it Kepler} satellite, Caceres et al. shows that the brightness fluctuations of the parent stars can be effectively reduced. Subsequent searches of transit-like signals from the ARIMA residuals resulted in a recovery of a significant fraction of confirmed exoplanets \cite{2019AJ....158...58C}.  

We started our experiment by testing whether the existing code {\tt auto.arima} from the {\tt R} package {\it forecast}\footnote{\href{https://www.r-project.org}{https://www.r-project.org}} is capable to fit the ARIMA model to GW data with the orders and the parameters of the model estimated automatically. {\tt auto.arima} was used in constructing ARPS pipeline \citep{2019AJ....158...58C}. 

We have tested whether {\tt auto.arima} can extract the waveform of GW150914 from LIGO data obtained by both detectors in Hanford (H) and Livingston (L) with a sampling frequency of 4~kHz. The data were downloaded from the event portal managed by the GW Open Science Center (GWOSC)\footnote{\href{https://www.gw-openscience.org/eventapi/html/allevents/}{https://www.gw-openscience.org/eventapi/html/allevents/}}.
However, since the strain amplitude is at an order of $h\lesssim 10^{-18}$, the modeling apparently suffered from an underflow problem. 

In order to circumvent the underflow, we have normalized the data to the order of unity.
In this case, {\tt auto.arima} yields the orders of (\textit{p,q,d})=(4,0,1). By subtracting the resultant model from the original data, we have obtained the residuals. However, even with the aid of a low-pass filter, we are unable to identify any waveform which resembles that of GW150914 from the residuals.
In view of such an undesirable behavior, instead of employing {\tt auto.arima} as in \cite{2019AJ....158...57C, 2019AJ....158...58C}, we are going to develop our own algorithm which is more suitable for reducing noise in GW data. 

\begin{figure}[htbp]
    \centering
    \includegraphics[width=0.6\linewidth]{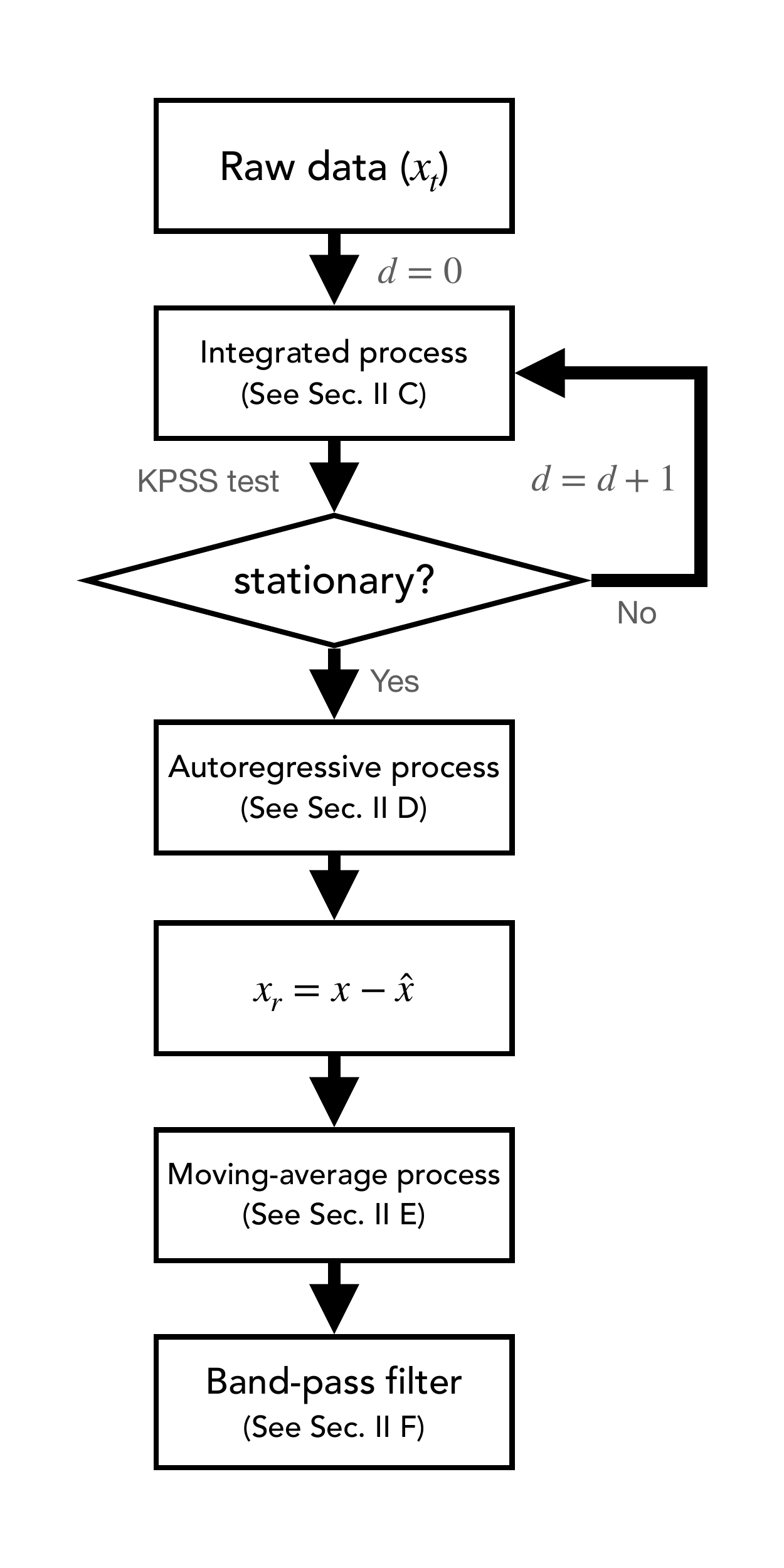}
    \caption{The structure of our proposed framework seqARIMA.}
    \label{fig:FlowChart}
\end{figure}
\begin{figure}
    \centering
    \includegraphics[width=\linewidth]{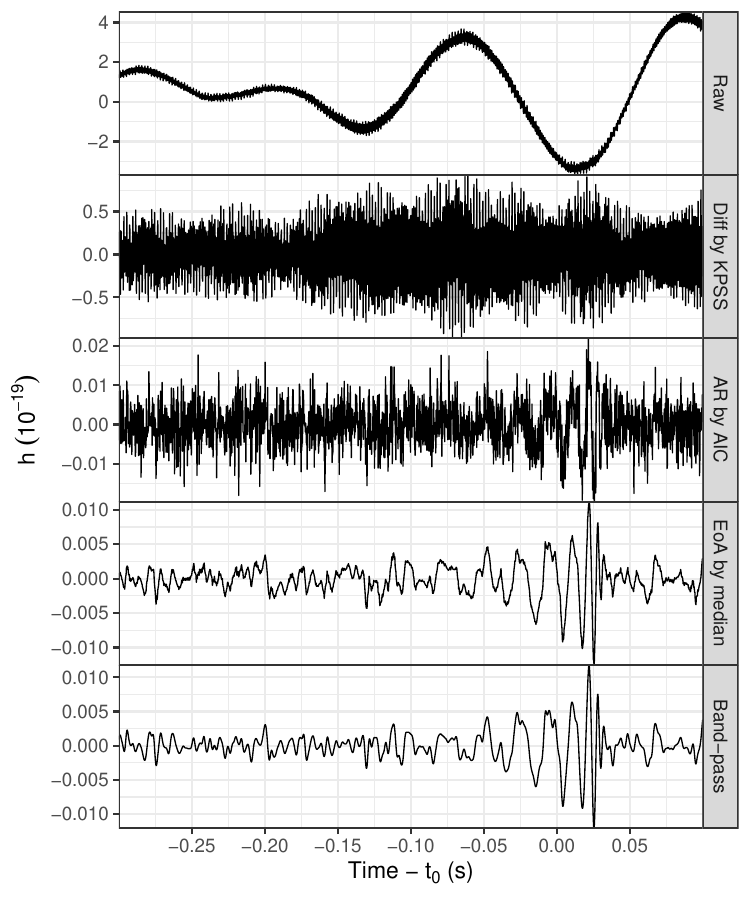}
    \caption{Illustrations for the effects of each stage in seqARIMA by using the LIGO-H data of GW150914.}
    \label{fig:ARIMAsteps}
\end{figure}

In our proposed framework, we break the noise reduction process  into a sequence of procedures as shown in \autoref{fig:FlowChart}. 
Hereafter we refer it as sequential ARIMA model (seqARIMA),
which consists of four stages:  integrated process,  autoregressive process, moving-average process, and bandpass filtering.

In this section, we take the LIGO data of GW150914 (Hanford, 32~s, 4~kHz sampling) to demonstrate the performance of seqARIMA. The effects of each stage in our procedure are illustrated in \autoref{fig:ARIMAsteps} and described in the following subsections (i.e. \autoref{sec:integProc}-\autoref{sec:bpProc}). 

\subsection{Integrated process}
\label{sec:integProc}

%

As a first step of our proposed framework, integrated process plays an essential role for ensuring the stationarity of a given GW time series.

For demonstrating the procedure, we start with the raw LIGO-H data of GW150914 (top panel in \autoref{fig:ARIMAsteps}) and with the parameter corresponding to the order of differencing initialized as {\tt d=0}. We employ the Kwiatkowski–Phillips–Schmidt–Shin (KPSS) test \citep{Kwiatkowski_Phillips_Schmidt_Shin_1992}, which is a standard test for stationarity \citep{FERRERPEREZ201728}, to examine whether the raw data exhibits trend and non-zero mean level. \autoref{fig:ARIMAsteps} clearly shows that the raw data is non-stationary.

Instead of ensuring global stationarity on the entire input time series, we consider local or segmented stationarity within time windows which are comparable to the signal duration from the typical BBH coalescence.
A segment within a time series refers to a sequence of data points collected or recorded at a given time interval.
Most time series segmentation algorithms can be classified into three primary categories: sliding windows, top-down, and bottom-up approaches \citep{keogh2004segmenting}.
Lovri\'{c} et al. have demonstrated that the process of segmenting time series into a limited number of homogeneous segments aids in the extraction of time segments with similar observations \cite{lovric2014algoritmic}.
These techniques process the input time series and return a piecewise linear representation (PLR).
Our method, however, divides time series into non-overlapping times series of equal length for the best performance.
Since the total length of the time series is much longer (32~s) than the duration of the typical BBH coalescence signal (i.e. $\lesssim$ 0.5 seconds), the whole time series is divided into 64 segments and the KPSS test is applied on each segment with the length 0.5~s (i.e. $t_{\rm s} = 0.5$).
If the $p$-values from KPSS test on all those segments are greater than or equal to our predefined threshold ($p$-value = 0.1), the given data is determined as a stationary time series. The threshold is chosen to be larger than the conventional value of $p$-value = 0.05 so as to reduce the false negatives.

Otherwise, if there is any segment exhibits non-stationary behavior, we modify the parameter $d$ as $d+1$ and apply differencing by \autoref{eq:integrate}. Such process will be iterated until $x_d$ satisfies the stationary condition by passing the KPSS test (See \autoref{alg:Integrated}). 
In the second panel of \autoref{fig:ARIMAsteps}, we show an optimal differencing model for our test data with $d=2$.

Since the result of hypothesis test can be influenced by the volume of data used, we further test the robustness of 
\autoref{alg:Integrated} by running KPSS tess on different $t_{s}$. In the aforementioned experiment, we took $t_{s}=0.5$~s which gives $\sim2000$ data points for a sampling rate of 4~kHz. For investigating the possible impact of non-stationarity detection by the length of data segment, we have re-run \autoref{alg:Integrated} on the same data by varying $t_{s}$ from 0.25~s to 0.75~s. And we found that all cases yield the same optimal differencing model with $d=2$. In view of this, we conclude that the results from KPSS test and hence \autoref{alg:Integrated} is robust and our adopted segment length of $t_{s}=0.5$~s is sufficient.


\begin{algorithm}[H]
	\SetKwInOut{Input}{Input}
	\SetKwInOut{Output}{Output}
        \SetKwInOut{Initialize}{Initialization}
	\caption{Integrated process (Sec. II C)}
    \label{alg:Integrated}

	\Input{$x_t$, the input time series, \\
		$t_{\rm t}$, the time length of $x_t$, \\
		$t_{\rm s}=0.5$, the length of each segment,\\
        $c_p = 0.1$, the threshold for the acceptance of p-values
	   	}
	\Output{$x_{d}$}
        \Initialize{$n_{\rm s} = \lceil t_{\rm t}/ t_{\rm s} \rceil$, the number of segments,\\
            $\mathbf{S} = \{ s_1, ..., s_{n_{\rm s}} \}$, the consecutive segments
	}

    $d = 0$\\
	\While{true}{
		$x_d = (1-B)^d x_t$
  
		\For{$i=1$ \KwTo $n_{\rm s}$}{
			Perform KPSS test on $s_{i}$ \\
			Compute the $p$-values for the level stationarity $p_L$-value and the trend stationarity $p_T$-value with $s_i$\\
            }
  
		\If{$p_{L} \ge c_p$ and $p_{T} \ge c_p$ $\forall s_{i \in \{1,2,...,n_s\}}$}{
            \Return $x_{d}$ \\
            } \Else{
                $d = d + 1$ \\
            }
	}

\end{algorithm}
\setcounter{AlgoLine}{0}

\subsection{Autoregressive process}
\label{sec:arProc}
Once the stationary data set is obtained as an output $x_d$ of the integrated process, we build the AR model of $x_d$. 
A set of AR models can be produced by:
\begin{equation}
    \hat{x}_{d, p} = \sum^{p}_{i=1} a_i x_{d,t-i} + \epsilon_t 
\end{equation}
where $p \in \{1,...,p_{\rm max}\}$.

For constructing a set of candidate models, we need to fix the upper-bound of $p$ which is set by the hyper-parameter $p_{\rm max}$. The optimal AR model is determined by model selection based on Akaike Information Criteria \citep[AIC]{Bozdogan1987}, which is defined as $AIC=2p + n\ln(\hat{\sigma_{p}}^2)$ where $n$ is the sample size and $\hat{\sigma_{p}}^2$ is the maximum likelihood estimator for the variance of the noise term. For each model in the set, AIC is calculated. And the model that attains the optimal AIC will be selected.

We have examined the effect with different values of $p_{\rm max}$ by subtracting the corresponding optimal AR model from the data (cf. \autoref{eq:resid}). The results are shown in \autoref{fig:ARs} which show that $p_{\rm max}=8192$ (corresponding to $\sim$ 2 seconds for 4 kHz sampling frequency) can lead to recognizable waveform in the residuals. 
We have examined whether the result can be further improved by setting $p_{\rm max}$ at higher values. However, among all the experiments presented in this work, we found that the optimal $p$ selected by AIC all converged below $8192$ even with $p_{\rm max}$ set at higher values. In view of this, we fixed $p_{\rm max}=8192$ as the hyper-parameter throughout this work for an efficient computation. 


In our framework, the model parameters (i.e. AR coefficients $\{a_{i}\}$) are estimated by Burg method, which fits the model to $x_d$ for minimizing the sums of squares of forward and backward linear prediction errors \citep{burg1968, kay1988modern}. The function {\tt ar.burg} from the {\tt R} package {\it stats} is adopted 
in our experiment.

\begin{figure*}[t!]
\captionsetup[subfigure]{labelformat=empty}
        \subfloat{
            \includegraphics[width=.32\linewidth]{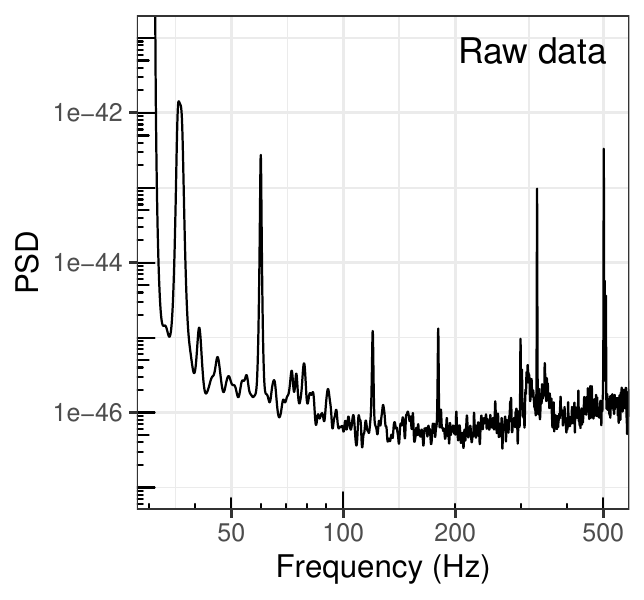}
        }\hfill
        \subfloat{
            \includegraphics[width=.32\linewidth]{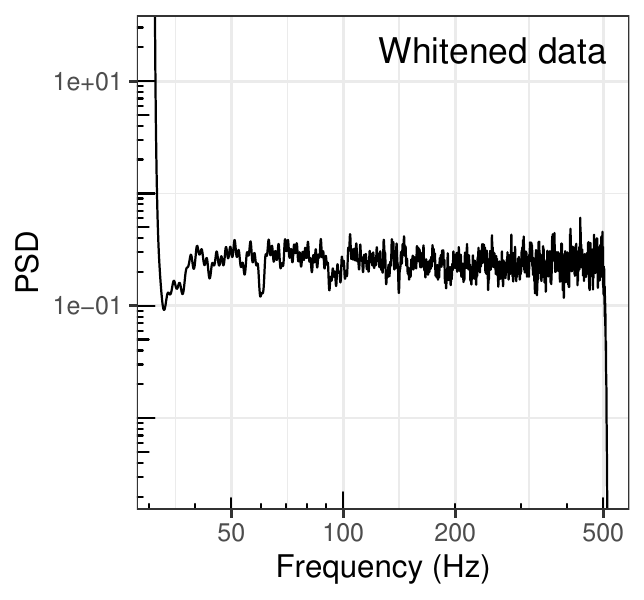}
        }\hfill
        \subfloat{
            \includegraphics[width=.32\linewidth]{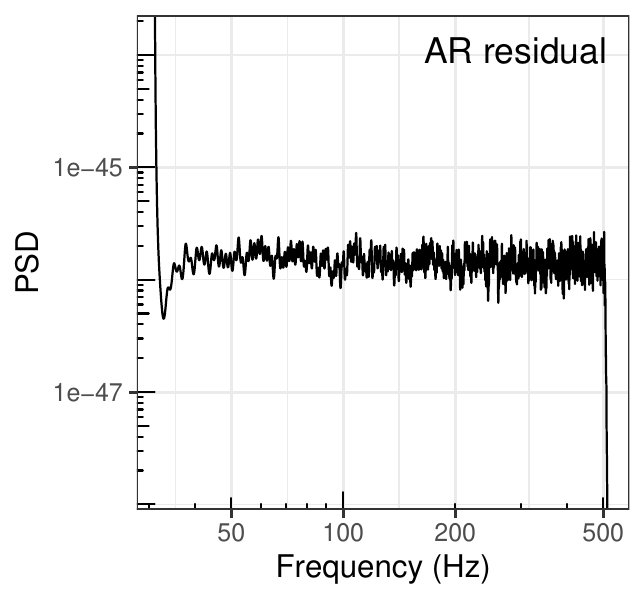}
        }\\
        \subfloat{
            \includegraphics[width=.325\linewidth]{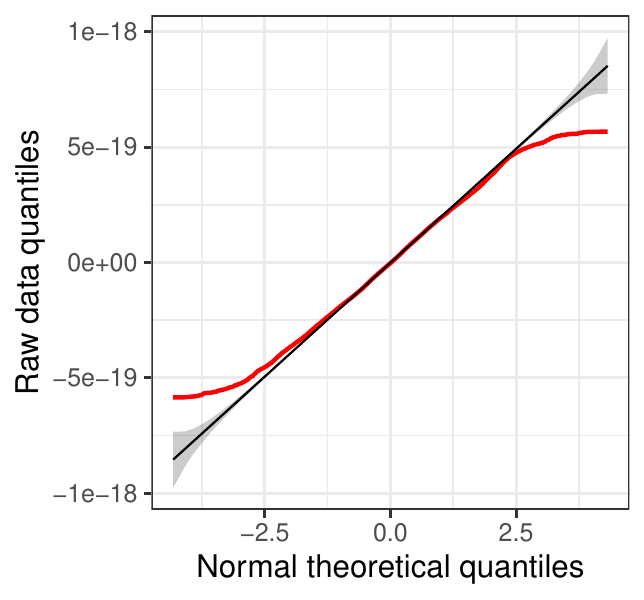}
        }\hfill
        \subfloat{
            \includegraphics[width=.325\linewidth]{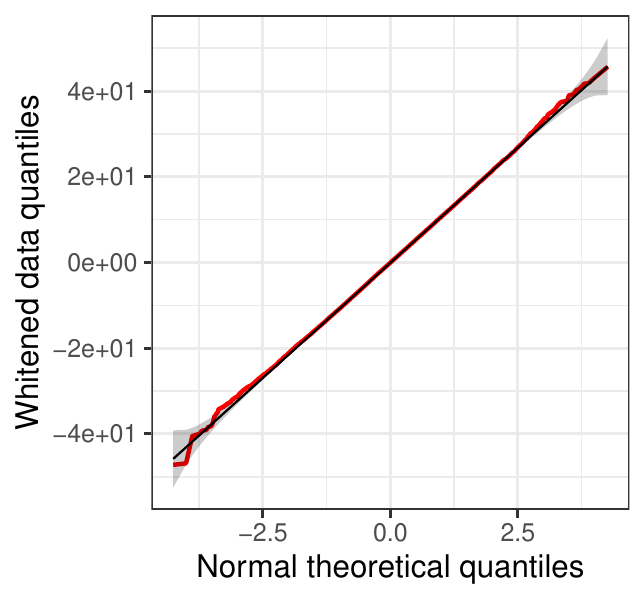}
        }\hfill
        \subfloat{
            \includegraphics[width=.325\linewidth]{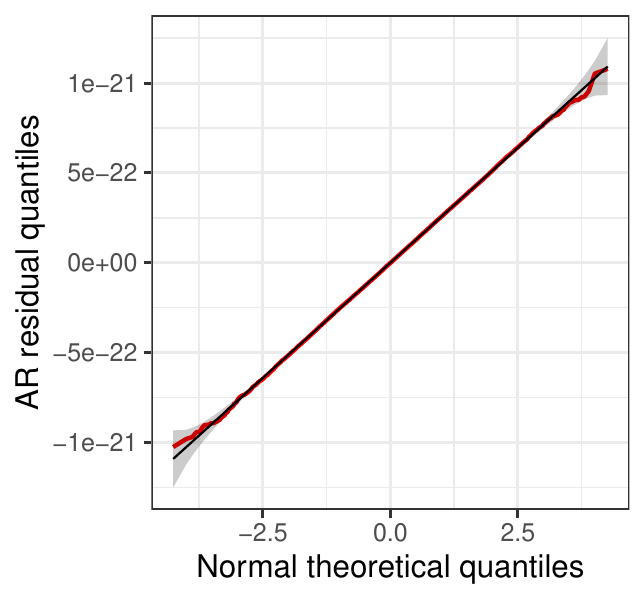}
        }
        
        \caption{({\it upper panels}) PSDs of the pure noise data, AR residuals and the whitened data. ({\it lower panels}) Q-Q plots for comparing the distributions of the raw/processed data (red lines) and with a Gaussian distribution (black lines).}
        \label{fig:ARI_psd_qq}
\end{figure*}

Since AR is the major component of our procedure, before we apply it to the data with a CBC signal embeded, we have first investigated its performance on the pure noise data and examined whether the processed data can satisfy the requirements of stationarity and normality \citep[cf.][]{PhysRevD.91.084034}. For this test, we have used both simulated and real noise data. We started by generating 100 simulated noise data of 32~s from sampling the updated Advanced LIGO sensitivity design curve \footnote{\href{https://dcc.ligo.org/LIGO-T1800044/public}{https://dcc.ligo.org/LIGO-T1800044/public}}. And we have processed them with \autoref{alg:Integrated} and \autoref{alg:arprocess}. For comparison, we have also separately processed the simulated noise with the standard whitening. To quantify the  difference between the distribution of the data from normality, we have run the Anderson-Darling (A-D) test \citep{10.1214/aoms/1177729437}. Taking the $p-$value of 0.05 as the benchmark for rejecting the null hypothesis, all the simulated data fail to pass the A-D test which yield a mean $p-$value of $\sim10^{-16}$. This suggests they are all significantly different from a Gaussian distribution. After subtracting the noise data from the AR models, $\sim70\%$ of these samples become conform with normality (yield a $p-$value $>0.05$). And we found that whitening results in a similar fraction that pass A-D test. 

On the other hand, all these simulated noise data are found to pass KPSS test and do not demonstrate any non-stationarity. In order to search for the non-stationary noise data for the experiment, we have searched over the LIGO data. We have chosen 24 time segments which do not encompass any confirmed GW events and yield a $p-$value smaller than 0.05 in the KPSS test. Also all these segments do not conform with normality which yield a mean $p-$value of $\sim10^{-14}$ in the A-D test. After whitening (or processing with our method), all 24 processed pure noise data pass the KPSS test (all yield $p-$value $>0.1$). Also, both methods result in similar fraction ($\sim65\%$) for passing the normality test. Therefore, we conclude that both our method and whitening have a comparable performance on the pure noise in attaining normality and stationarity. As an example, we compare the power spectral density (PSD) for one of our real noise sample with those of AR residuals and whitened data in \autoref{fig:ARI_psd_qq}. These plots also demonstrate the capability of a AR model in line removal. In the low panels, we have also constructed the quantile-quantile (Q-Q) plots for comparing the distribution of the data with the normal distribution. They clearly show that both AR residuals and the whitened data distribute as a Gaussian. \\

For our test data which encompasses the transient signal from GW150914, an optimal AR order of $p=7931$ is obtained. Before passing to the next stage of processing, we obtain the residual time series $x_{r}$ by subtracting the optimal model from $x_d$ (cf. \autoref{eq:resid}, \autoref{alg:arprocess}). 
The waveform of AR residual data is shown in the third panel in \autoref{fig:ARIMAsteps}, in which the modulation resulting from the BBH coalescence starts emerging. To further suppress the random fluctuation in $x_{r}$,  we proceed to the next stage (see below). 


\begin{algorithm}[H]
	\SetKwInOut{Input}{Input}
	\SetKwInOut{Output}{Output}
	\caption{Autoregressive process (Sec. II D)}
    \label{alg:arprocess}
    
	\Input{$x_d$, the input time series, \\
		   $p_{\rm max} = 8192$: the maximum number of the order of the autoregressive process \\}
	\Output{$x_{r}$}

            Estimate the parameters $\phi_p$ of an AR($p$) model on $x_d$ with the variance $\sigma^2_p$ calculated using Burg's method and \\

    $\{\phi_p, p_{\rm opt}\} = \mathop{\arg\min}\limits_{\phi_p, p_{\rm opt}}{AIC}$ for $p \in \{1, ..., p_{\rm max}\}$ \\    
    with $AIC = 2p + n\ln(\hat{\sigma_{p}}^2)$ \\

 
	Obtain time series $\hat{x}_{p_{\rm opt}}$ from the optimal model $\{\phi_p, p_{\rm opt}\}$\\
 
	\Return the residual $x_{r} = x_d - \hat{x}_{p_{\rm opt}}$ \\
\end{algorithm}
\setcounter{AlgoLine}{0}


\begin{figure}[htbp]
    \centering
    \includegraphics[width=\linewidth]{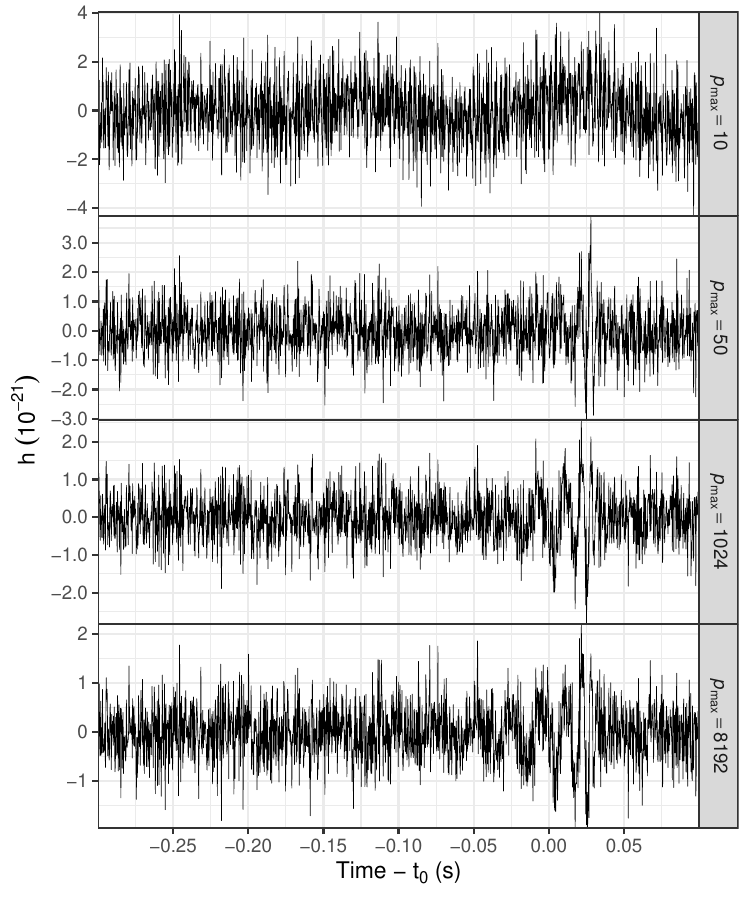}
    \caption{Effects of AR with different $p_{\rm max}$ on the difference data of $d=2$. These time series are the residuals obtained by subtracting the fitted AR model from the data (i.e. $x_{r}$ outputted by \autoref{alg:arprocess}).}
    \label{fig:ARs}
\end{figure}

\subsection{Moving-average process}
\label{sec:maProc}

Different from the conventional ARIMA model (\autoref{eq:ARIMA}) in which MA performs the regression with the past forecast errors $\epsilon_{i}$. In our framework, MA refers to the method of estimating the trend in the residuals which is taken as a form of low-pass finite impulse response filter. The process is expressed as follows:

\begin{equation}
    x_{\rm ma} = \frac{1}{q} \sum_{j=-k}^{k} x_{t+j}
    \label{eq:MAsmoother}
\end{equation}
where $q$ is the order of MA and $k=(q-1)/2$.
Since we consider a two-sided (centered) MA, if an even value of order $q$ is specified, two MAs of $k$ rounded-down and rounded-up will be averaged \citep[cf.][]{hyndman2014forecasting}. 

\begin{figure}[htbp]
    \centering
    \includegraphics[width=\linewidth]{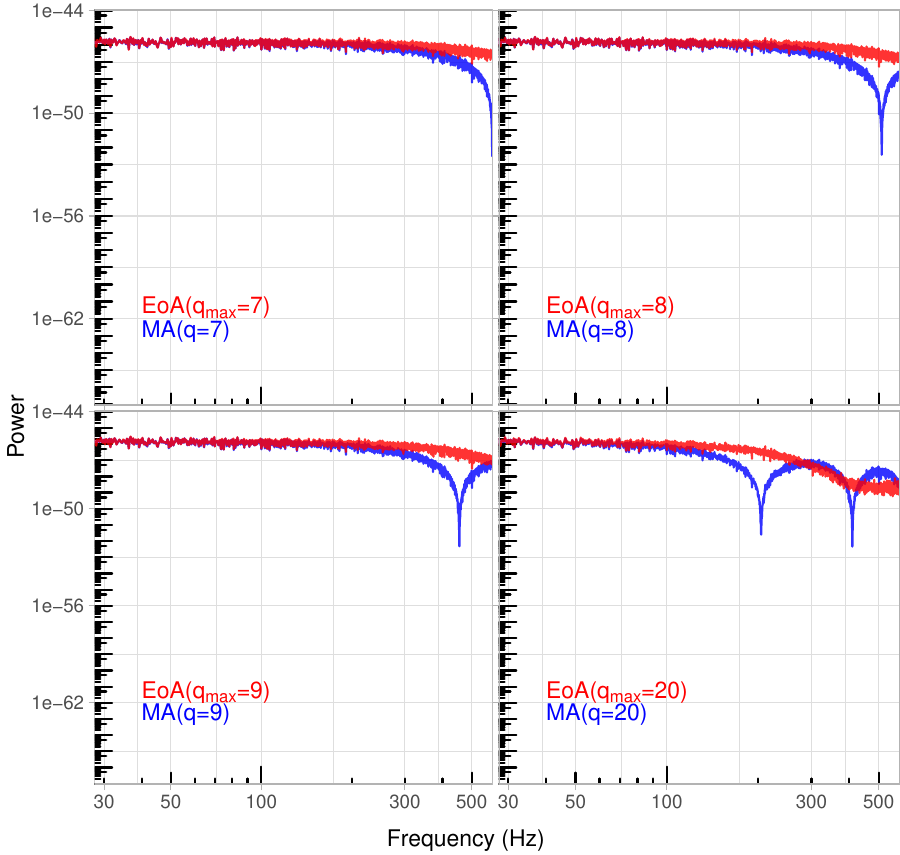}
    \caption{PSDs of the outputs from EoA with different $q_{\rm max}$ (red lines). These are compared with the corresponding result from MA with single value of $q$ (blue lines). Cases of different $q_{\rm max}$ and $q$ are shown.}
    \label{fig:MAsPSD}
\end{figure}

For choosing the value of $q$, a model with small $q$ might have the signal remain buried by the random fluctuations in $x_{r}$. On the other hand, a large $q$ can smear out the signal. Since the GW from the BBH coalescence has its frequency varying, a MA model with a fixed $q$ can suffer from the aforementioned trade-off. Another problem of a single MA model is found when large values of $q$ are adopted.
In the \autoref{fig:MAsPSD}, we show the power spectral density (PSD) of the output from the MA model with different $q$ annotated as blue lines in each panel. Within our concerned frequency band (32$-$512~Hz), 
We found that power spectral leakage starts appearing with $q\geq7$, which can be possibly resulted from over-smoothing.

To overcome the aforementioned problems, rather than using only a single MA($q$), we adopted the method of Ensemble of Averages (EoA) \citep[cf.][]{arpit2022ensemble} which combines MAs from a range of $q$ ($q\in\{$1...$q_{\rm max}\}$). It aggregates a number of MAs with an ensemble of moving averages. 
In our work, we utilize EoA and demonstrate that using median as the collector function can be a very effective filter (\autoref{alg:macprocess}).

\begin{figure}[htbp]
    \centering
    \includegraphics[width=\linewidth]{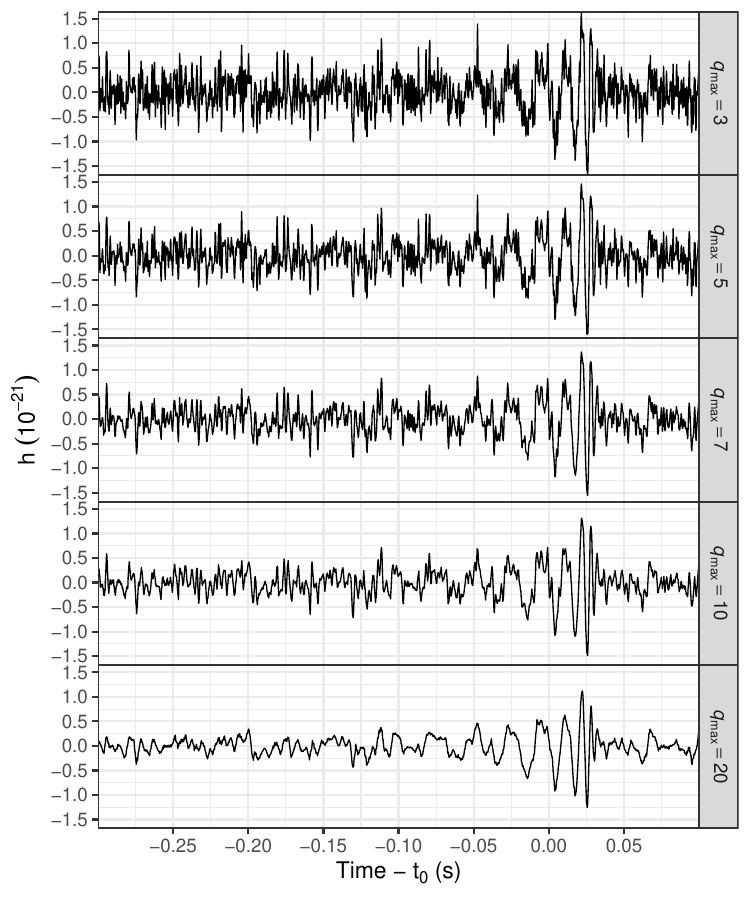}
    \caption{EoA with a median collector on the output of AR stage. Collector function aggregates MAs from $q=1$ to  $q=q_{\rm max}$.}
    \label{fig:MACs_med}
\end{figure}

\autoref{fig:MACs_med} shows the outputs of EoA for different choices of $q_{\rm max}$. Empirically, we found that $q_{\rm max}=20$ with a median collector function gives a desirable result. It can eliminate most random fluctuations and retains the signal fidelity as all three stages of coalescence (i.e., inspiral, merger, and ringdown) can be clearly visualized. Furthermore, with EoA, the problem of spectral power leakage in PSD is resolved (annotated as red lines in \autoref{fig:MAsPSD}).

\begin{algorithm}[H]
	\SetKwInOut{Input}{Input}
	\SetKwInOut{Output}{Output}
	\caption{Moving-average process (Sec. II E)}
    \label{alg:macprocess}
    
	\Input{$x_{r}$, \\
 		   $ q\in \{q_{\rm min}, ..., q_{\rm max}\}$, the order of the moving average\\ where $q_{\rm min}=1$ and $q_{\rm max} = 20$ by default
      }
	\Output{$x_{\rm EoA}$}
 
	\For{$q = q_{\rm min}$ \KwTo $q_{\rm max}$}{
		$x_{q} = \frac{1}{q} \sum^k_{j=-k} x_{r+j}$
	}
        $\hat{x}=median( \{x_{q_{\rm min}}, ..., x_{q_{\rm max}} \})$\\
	\Return $\hat{x}$ \\
\end{algorithm}

\subsection{Bandpass filtering}
\label{sec:bpProc}
In the last step of our framework, we have $x_{\rm EoA}$ bandpass filtered in the frequency range of $32-512$~Hz for removing noise out of this band (e.g. seismic noise at low frequencies and photon shot noise at high frequencies). We adopt the Finite Impulse Response (FIR) filter by using the functions {\tt filtfilt} and {\tt fir1} from the {\tt R} package {\it signal}\footnote{\href{https://cran.r-project.org/web/packages/signal/index.html}{https://cran.r-project.org/web/packages/signal/index.html}}.  
The bandpass filtered signal of GW150914 is shown at the bottom panel in \autoref{fig:ARIMAsteps}. 

In comparison with the output from the previous step (i.e. $x_{\rm EoA}$), no significant improvement can be found as a result of bandpass, which indicates that seqARIMA has already efficiently suppressed the noise in the raw data.

Although the bandpass does not appear to be necessary in the case of GW150914, we keep it in our framework to ensure all unwanted modulations outside this band are removed for the sake of comparing with the whitening results.

 \section{Experimental results}

 \subsection{Simulated data}

For comparing the performance between the frequency-domain whitening filter and seqARIMA in noise reduction as well as waveform visualization, we have carried out a series of experiments. We started by simulating clean waveforms of a BBH coalescence at different luminosity distance $d_{L}$ by the code {\tt get\_td\_waveform} from {\it pycbc} \footnote{\href{https://pycbc.org}{https://pycbc.org}} with the model {\tt SEOBNRv4\_opt}.


We have considered $d_L$ in a range from 200-4000 Mpc with a step size of $\Delta d_L$=200 Mpc. 
In order to analyse the denoising performance for a variety of waveform, for each $d_{L}$, we have generated 100 waveform of randomly sampled individual component masses $m_{1}$ and $m_{2}$.
For the other parameters such as dimensionless spin and eccentricity, the default values of {\tt get\_td\_waveform} are adopted (See \footnote{\href{https://pycbc.org/pycbc/latest/html/pycbc.waveform.html}{https://pycbc.org/pycbc/latest/html/pycbc.waveform.html}}).
These waveform are defined as the signals $s$.

For the sampling of waveform parameters, we have firstly fitted the distributions of  $m_{1}$ and $m_{2}$ from all the 81 confirmed BBH CBC events with the {\tt R} package {\it gamlss}\footnote{\href{https://www.gamlss.com}{https://www.gamlss.com}}. Among all the distribution functions available in {\it gamlss}\footnote{\href{https://search.r-project.org/CRAN/refmans/gamlss.dist/html/gamlss.family.html}{https://search.r-project.org/CRAN/refmans/gamlss.dist/html/gamlss.family.html}}, generalized Beta distributions of second kind provides the best description in accordance with AIC. And we sampled $m_{1}$ and $m_{2}$ from these best-fitted distributions. 



For $d_L$, we have generated 100 noise data of 32~s, which is defined as $n$. They are sampled from a PSD simulated by {\tt aLIGOZeroDetHighPower} in {\it pycbc} from {\tt LALsimulation} with {\tt low\_freq\_cutoff} of 15 Hz. Each of them are generated with different random seed. 
The preparation of the simulated data was finished by injecting $s$ in a random time location of $n$.

This simulated dataset allows us to compare the performance of seqARIMA and whitening filter in extracting the injected signal at varying $d_L$.
In both methods of seqARIMA and whitening, the same bandpass filter of $32-512$~Hz were applied. 
For whitening, we have adopted a segment length of 4~s and the an overlap percentage of $50\%$ in all experiments. Such choices of whitening parameters follow the standards given by the {\it pycbc} documentation \footnote{\href{https://pycbc.org}{https://pycbc.org}}

For quantifying the fidelity of the extracted signal, we computed the cross-correlation functions (CCFs) defined as,
\begin{equation}
    {\rm CCF}(t') = \sum_{t=-\infty}^{\infty} s(t) \hat{s}(t-t')
    \label{eq:CCF}
\end{equation}
where $s$ is the simulated waveform and $\hat{s}$ is the denoised data. Then we obtained the maximum values of $|{\rm CCF}|$, ${\rm CCF_{max}}$, as the metric of measuring the similarity between the $s$ and $\hat{s}$. 
In order to evaluate the noise reduction performance, we also computed the root-mean-square errors (RMSEs) defined as,
\begin{equation}
    {\rm RMSE} = \sqrt{\frac{\sum^{N}(s-\hat{s})^2}{N}}
    \label{eq:RMSE}
\end{equation}
where $N$ is the length of data, which reflects how the noise is suppressed in the whole time series.
For each $d_L$, we have re-sampled $n$ with 100 different random seeds and computed the median and the 95\% confidence interval of ${\rm CCF}_{\rm max}$ and RMSE from this sample.

\begin{figure}[htbp]
    \centering
    \includegraphics[width=\linewidth]{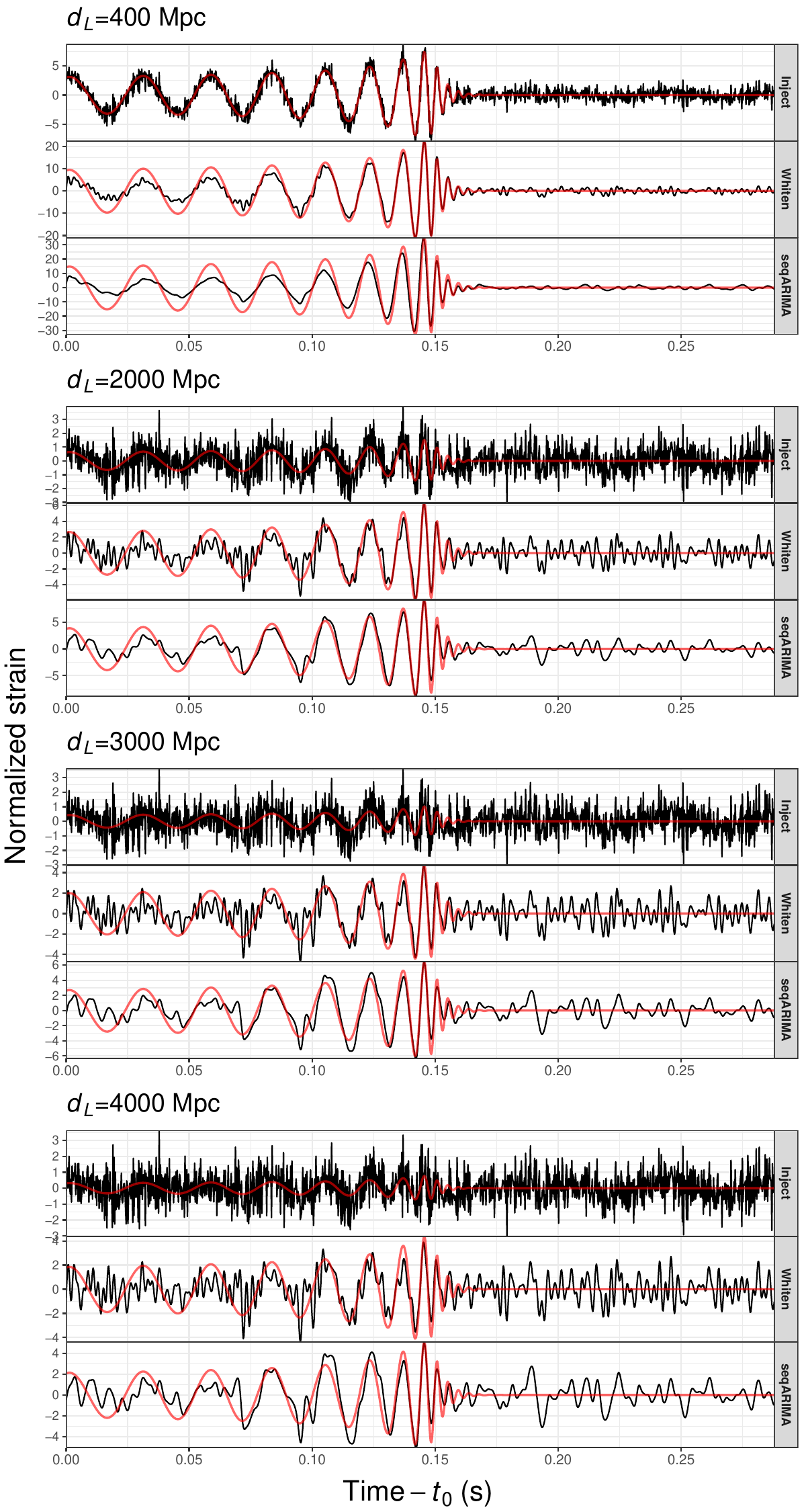}

     \caption{The comparison of extracted waveforms from the simulated data ({\it top panel}) by whitening ({\it middle panel}) and seqARIMA denoising ({\it bottom panel}) for $d_L=$400, 2000, 3000, and 4000 Mpc, from {\it top} to {\it bottom}, respectively. In each panel, we have overlaid the injected signal (red lines) on the data (black lines).  For the sake of comparison, the strain amplitudes are normalized in each panel.}
     \label{fig:SimBetas}
\end{figure}

In \autoref{fig:SimBetas}, the results are shown for $d_L=$400, 2000, 3000, and 4000 Mpc. 
For a visual comparison of the similarity of the extracted signal and the 
injected waveform $s$, we have also overlaid $s$ in all the panels of \autoref{fig:SimBetas} as the red solid curves. 

\begin{figure}[htbp]
\begin{center}
    \includegraphics[width=\linewidth]{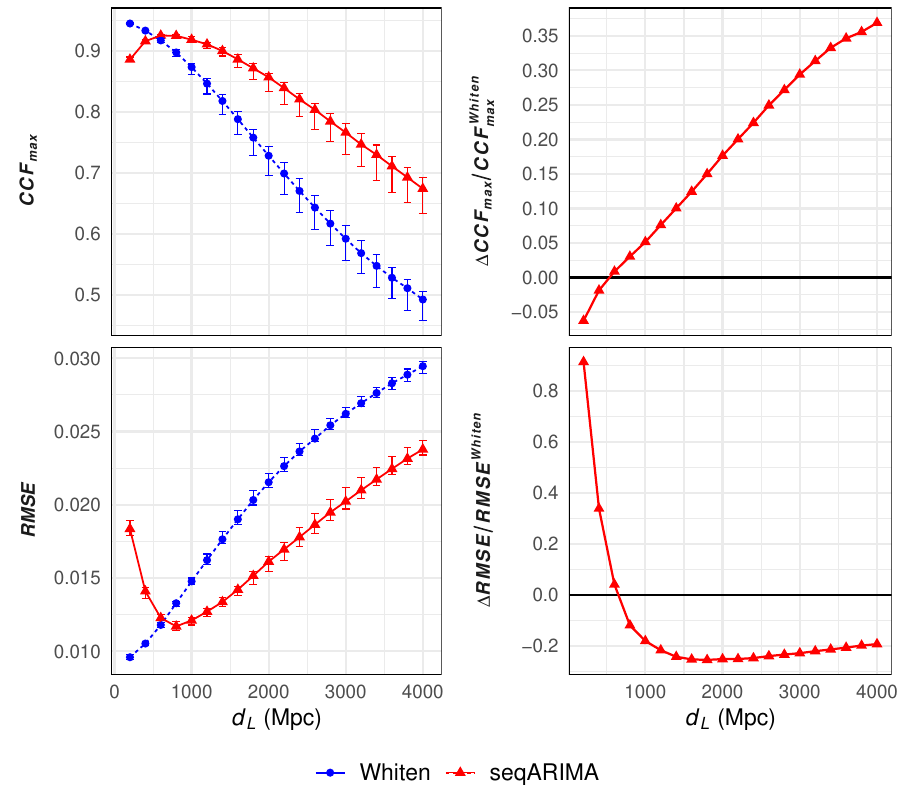}
\end{center}

    \caption{({\it left panel}) The comparison of ${\rm CCF_{max}}$ and RMSE between the injected signal and the waveform extracted from seqARIMA (red triangles) and the whitening process (blue circles) with varying $d_L$. The error bars represent 95\% confidence errors calculated from 100 sampled parameters and 100 different random seeds. ({\it right panel}) Fractional difference of ${\rm CCF_{max}}$ and RMSE resulted from seqARIMA denoising with respect to the corresponding metrics resulted from whitening as a function of $d_L$.}
     \label{fig:SimCCF}
\end{figure}

In the left panels of \autoref{fig:SimCCF}, we compare how ${\rm CCF}_{\rm max}$ and RMSE vary with $d_L$ in both schemes. The error bars represent 95\% confidence intervals calculated from 100 simulated waveform with randomly sampled
$m_1$ and $m_2$ as well as different random seeds for generating the noise. Comparing the extracted signals by these two methods, we found that those obtained by seqARIMA generally have a larger degree of similarity with $s$ and lower level of noise. Although whitening process attains better results for small distance ($d_{L}\leq600$~Mpc), seqARIMA has shown advantage in de-noising for increasing $d_L$  (i.e. larger ${\rm CCF}_{\rm max}$ and reduced RMSE).


 In the right panels of \autoref{fig:SimCCF}, we show the fractional improvements in both metrics as yielded by seqARIMA at different $d_L$. 
Comparing with the whitening results at $d_L=4000$ Mpc, seqARIMA has improved ${\rm CCF}_{\rm max}$ by $\sim42~$\% and suppressed RMSE by $\sim 23~$\%.




\subsection{LIGO data}
To demonstrate the capability of seqARIMA in handling real data, we have attempted to extract the signals from a number of known GW events from the LIGO data. In this test, we have chosen all the events in GWTC-1 as observed during the first and second observation runs (O1 and O2) in 2015-2017 \citep{PhysRevX.9.031040} plus two additional interesting events. All the data with a length of 4096~s with 4~kHz sampling frequency are obtained from GWOSC. Except for the NS-NS merger GW170817, we windowed the 4096~s data with a frame of 32~s for all the events. For GW170817, because of its much longer timescale, we apply a window of 50~s instead.

\begin{table}[htbp]
\centering
\resizebox{\columnwidth}{!}{%
    \begin{tabular}{l|c|cc|cc}
    \hline
\multirow{2}{*}{Event Name}       & \multirow{2}{*}{Dataset}   & \multicolumn{2}{c|}{RMSE}      & \multicolumn{2}{c}{CCF$_{\rm max}$}        \\ 
                                                               \cline{3-6}
                                  &                            & H1       & L1                  & H1       & L1                       \\ \hline
\multirow{3}{*}{GW150914}         & Whiten                     & 0.034    & 0.0288              & 0.628    & 0.488                    \\ 
                                  & seqARIMA                   & 0.024    & 0.0321              & 0.646    & 0.607                    \\ 
                                  &                      & -29.4 \% & +11.4 \%            & +2.83 \% & +24.4 \%                 \\ \hline
\multirow{3}{*}{GW151012}         & Whiten                     & 0.0318   & 0.0326              & 0.127    & 0.0922                   \\ 
                                  & seqARIMA                   & 0.0304   & 0.0311              & 0.201    & 0.177                    \\ 
                                  &                      & -4.39 \% & -4.71 \%            & +57.9 \% & +92.1 \%                 \\ \hline
\multirow{3}{*}{GW151226}         & Whiten                     & 0.0119   & 0.0119              & 0.0763   & 0.0736                    \\ 
                                  & seqARIMA                   & 0.0114   & 0.0117              & 0.145    & 0.109                    \\ 
                                  &                      & -4.05 \% & -1.81 \%            & +90.6 \% & +47.6 \%                 \\ \hline
\multirow{3}{*}{GW170104}         & Whiten                     & 0.0289   & 0.0283              & 0.212    & 0.212                    \\ 
                                  & seqARIMA                   & 0.0262   & 0.0264              & 0.298    & 0.293                    \\ 
                                  &                      & -9.27 \% & -6.71 \%            & +40.3 \% & +38.2 \%                 \\ \hline
\multirow{3}{*}{GW170608}         & Whiten                     & 0.0145   & 0.0144              & 0.0926   & 0.105                    \\ 
                                  & seqARIMA                   & 0.0138   & 0.0143              & 0.18     & 0.143                    \\ 
                                  &                      & -4.92 \% & -1.23 \%            & +94 \%   & +37 \%                   \\ \hline
\multirow{3}{*}{GW170729}         & Whiten                     & 0.0375   & 0.0391              & 0.278    & 0.379                    \\ 
                                  & seqARIMA                   & 0.033    & 0.0309              & 0.443    & 0.51                     \\ 
                                  &                      & -12.2 \% & -20.9 \%            & +59.7 \% & +34.5 \%                 \\ \hline
\multirow{3}{*}{GW170809}         & Whiten                     & 0.0328   & 0.0309              & 0.246    & 0.305                    \\ 
                                  & seqARIMA                   & 0.0315   & 0.0304              & 0.367    & 0.35                     \\ 
                                  &                      & -3.75\%  & -1.5 \%             & +49.4 \% & +15 \%                   \\ \hline
\multirow{3}{*}{GW170814}         & Whiten                     & 0.0309   & 0.027               & 0.303    & 0.402                    \\ 
                                  & seqARIMA                   & 0.0255   & 0.0276              & 0.467    & 0.546                    \\ 
                                  &                      & -17.6 \% & +2.0 \%             & +54 \%   & +35.8 \%                 \\ \hline
\multirow{3}{*}{GW170817}         & Whiten                     & 0.00836  & 0.00828             & 0.0719   & 0.106                    \\ 
                                  & seqARIMA                   & 0.00816  & 0.00793             & 0.113    & 0.163                    \\ 
                                  &                      & -2.38 \% & -4.16 \%            & +57.1 \% & +54.1 \%                 \\ \hline
\multirow{3}{*}{GW170818}         & Whiten                     & 0.035    & 0.0335              & 0.119    & 0.273                    \\ 
                                  & seqARIMA                   & 0.0325   & 0.0319              & 0.218    & 0.359                    \\ 
                                  &                      & -7.19 \% & -4.99 \%            & +82.9 \% & +31.2 \%                 \\ \hline
\multirow{3}{*}{GW170823}         & Whiten                     & 0.0336   & 0.0349              & 0.306    & 0.3                      \\ 
                                  & seqARIMA                   & 0.0352   & 0.0337              & 0.434    & 0.362                    \\ 
                                  &                      & +4.78 \% & -3.51 \%            & +41.7 \% & +20.7 \%                 \\ \hline
\multirow{3}{*}{GW190814}         & Whiten                     & 0.0118   & 0.0117              & 0.0951   & 0.109                    \\ 
                                  & seqARIMA                   & 0.0113   & 0.0112              & 0.171    & 0.182                    \\ 
                                  &                      & -4.68 \% & -4.62 \%            & +79.7 \% & +67.2 \%                 \\ \hline
\multirow{3}{*}{GW200105} & Whiten                     & -        & 0.0121              & -        & 0.0316                   \\ 
                                  & seqARIMA                   & -        & 0.0115              & -        & 0.174                    \\ 
                                  &                      & -        & -5.59 \%            & -        & +449 \%                  \\ \hline
    \end{tabular}
}
    \caption{The comparison of RMSE and CCF$_{\rm max}$ yielded by whitening and seqARIMA on 13 confirmed CBC events as observed by LIGO with reference to the model waveform as specified in the literature. The third row of each event show the percentage change resulted from seqARIMA with respect to whitening. For GW200105, LIGO-H was not operational during this event as hence there is no data available \cite{2021ApJ...915L...5A}.}
\label{tab: RealData_Metrics}
\end{table}

\subsubsection{GWTC-1 events}
\begin{figure*}[htbp]
     \centering
     \begin{subfigure}[b]{0.45\linewidth}
         \centering
         \includegraphics[width=\linewidth]{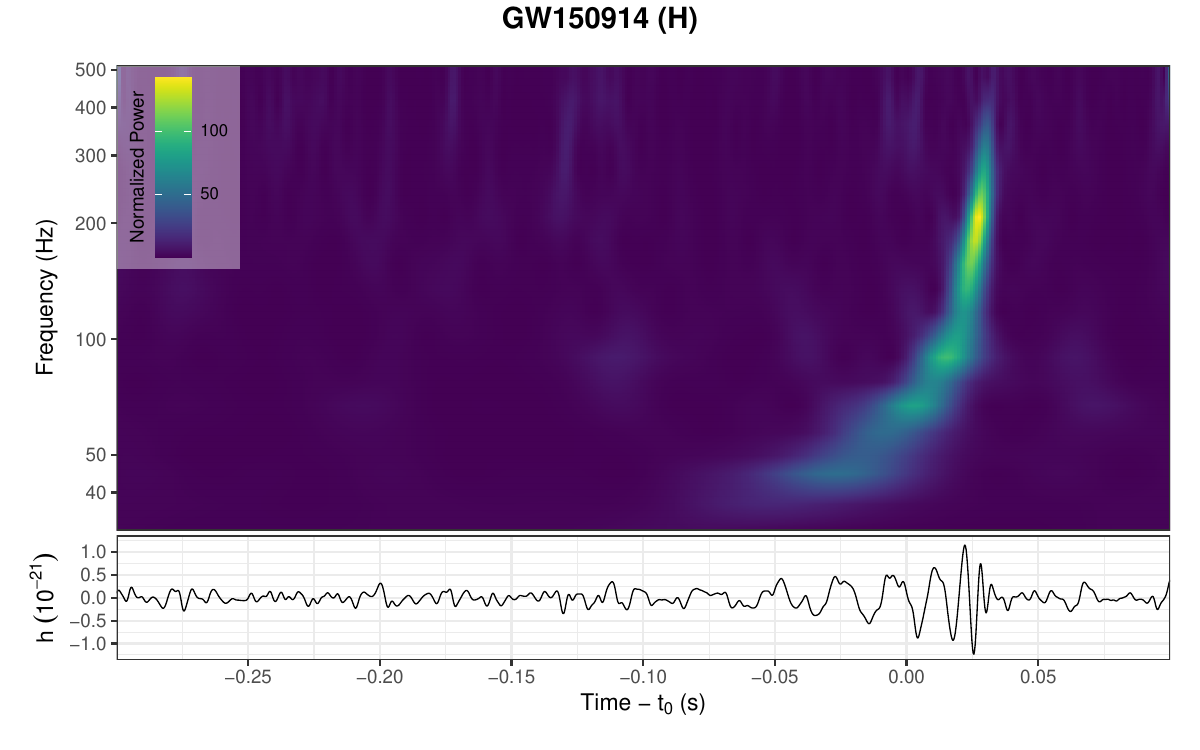}
     \end{subfigure}
     \begin{subfigure}[b]{0.45\linewidth}
         \centering
         \includegraphics[width=\linewidth]{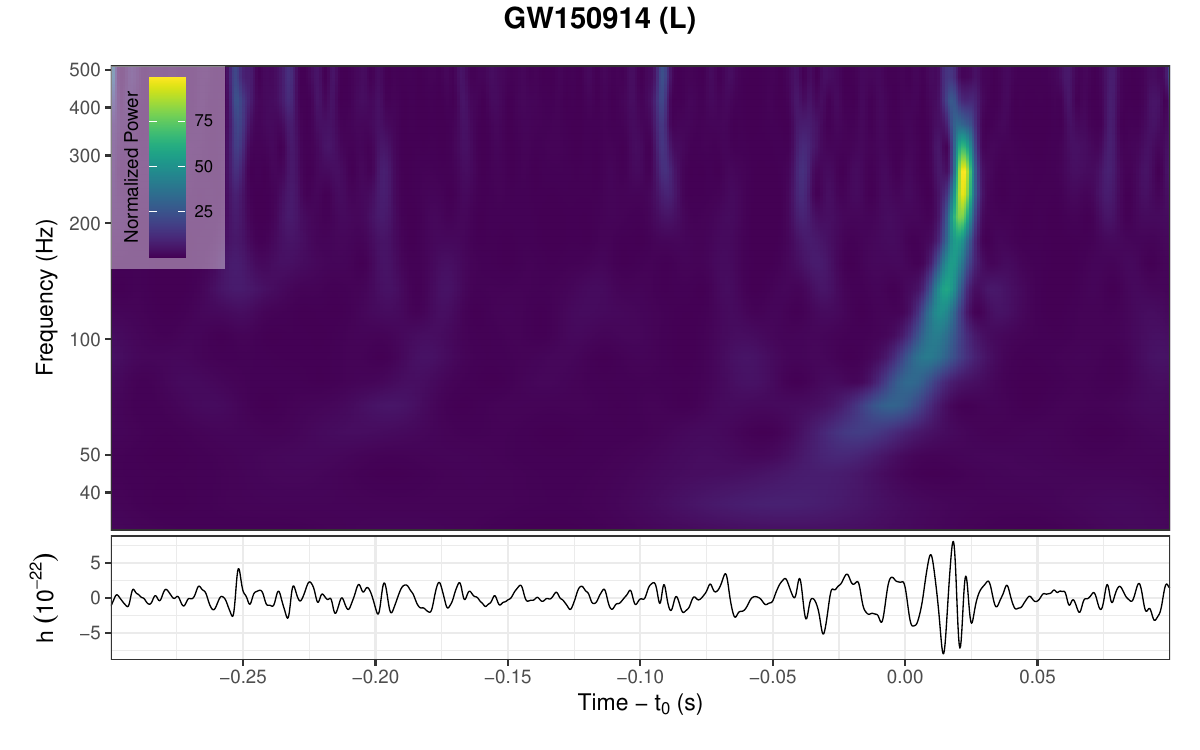}
     \end{subfigure}

     \begin{subfigure}[b]{0.45\linewidth}
         \centering
         \includegraphics[width=\linewidth]{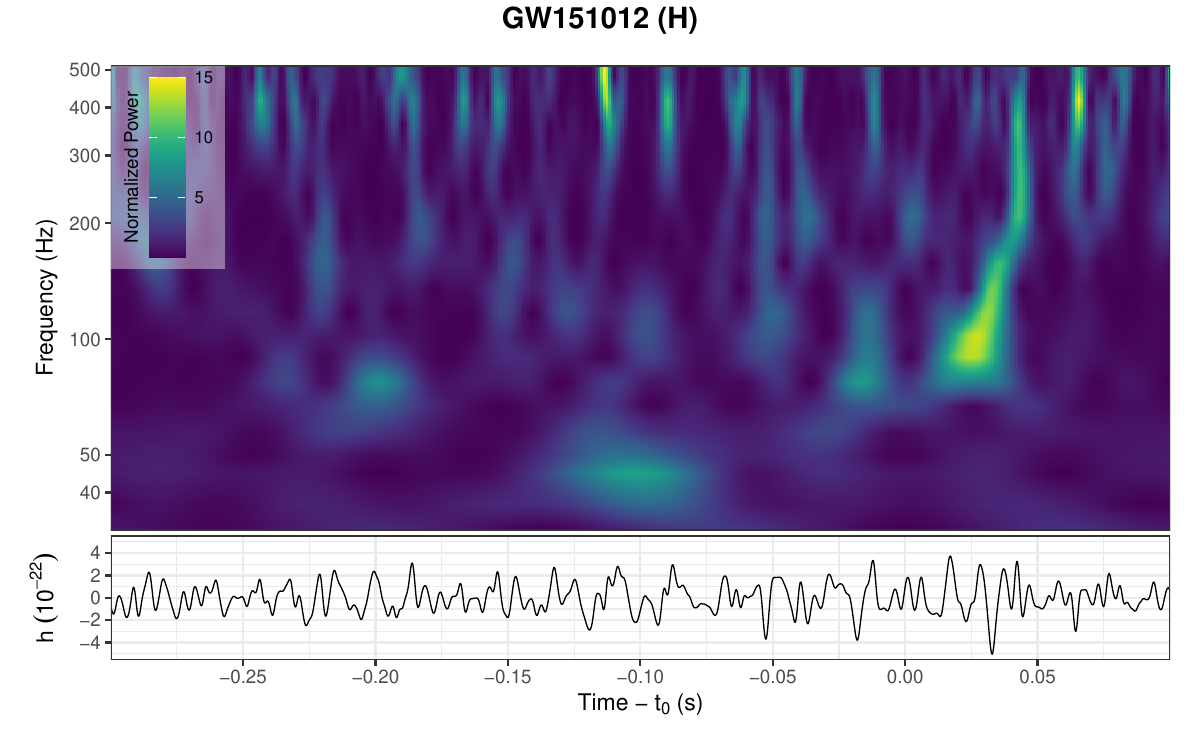}
     \end{subfigure}
     \begin{subfigure}[b]{0.45\linewidth}
         \centering
         \includegraphics[width=\linewidth]{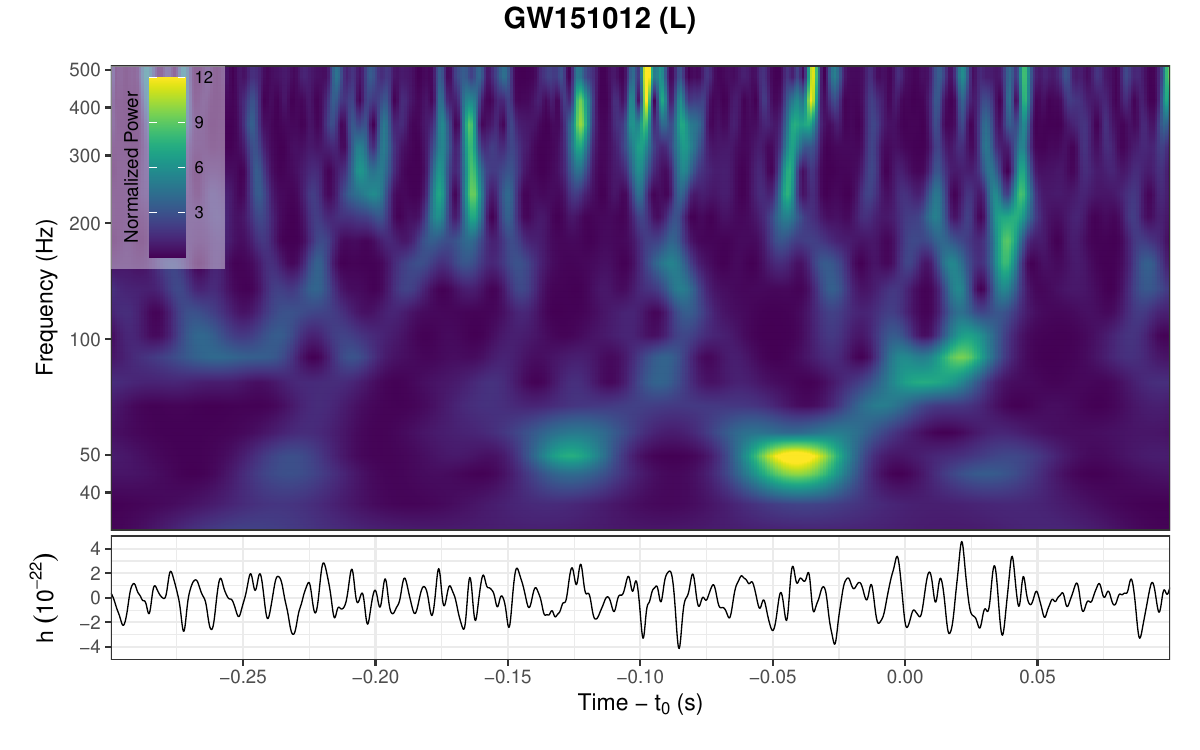}
     \end{subfigure}
     
     \begin{subfigure}[b]{0.45\linewidth}
         \centering
         \includegraphics[width=\linewidth]{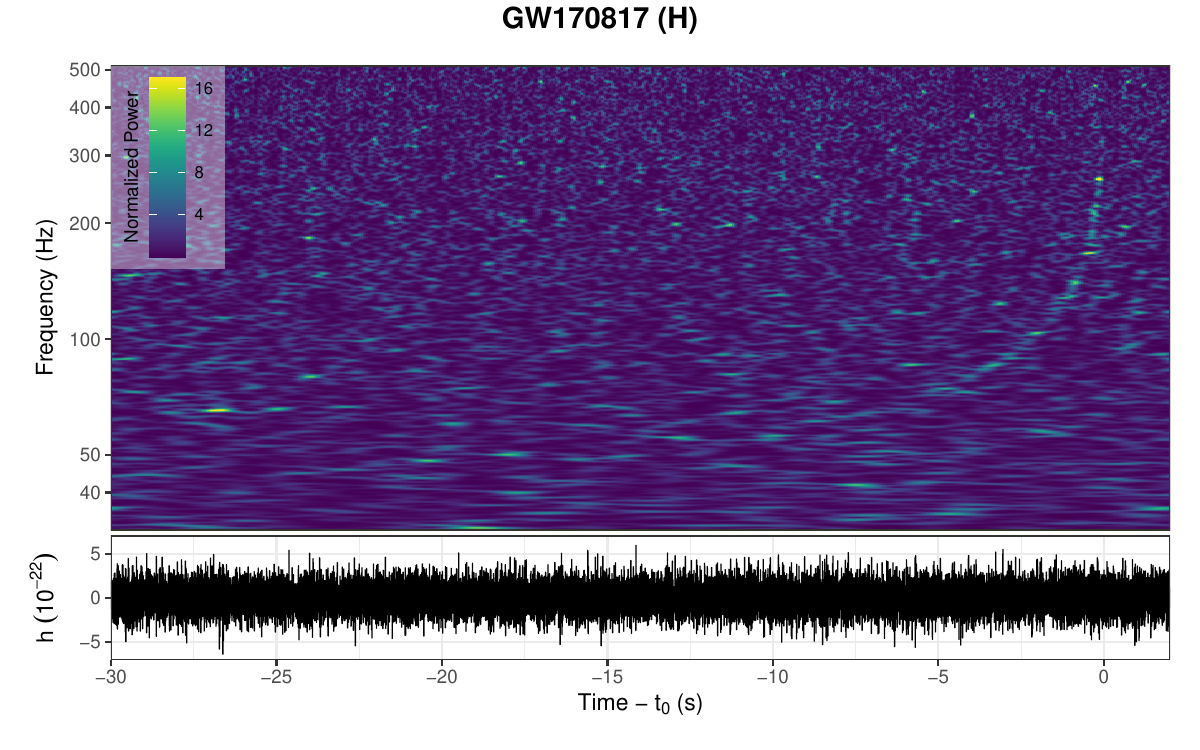}
     \end{subfigure}
     \begin{subfigure}[b]{0.45\linewidth}
         \centering
         \includegraphics[width=\linewidth]{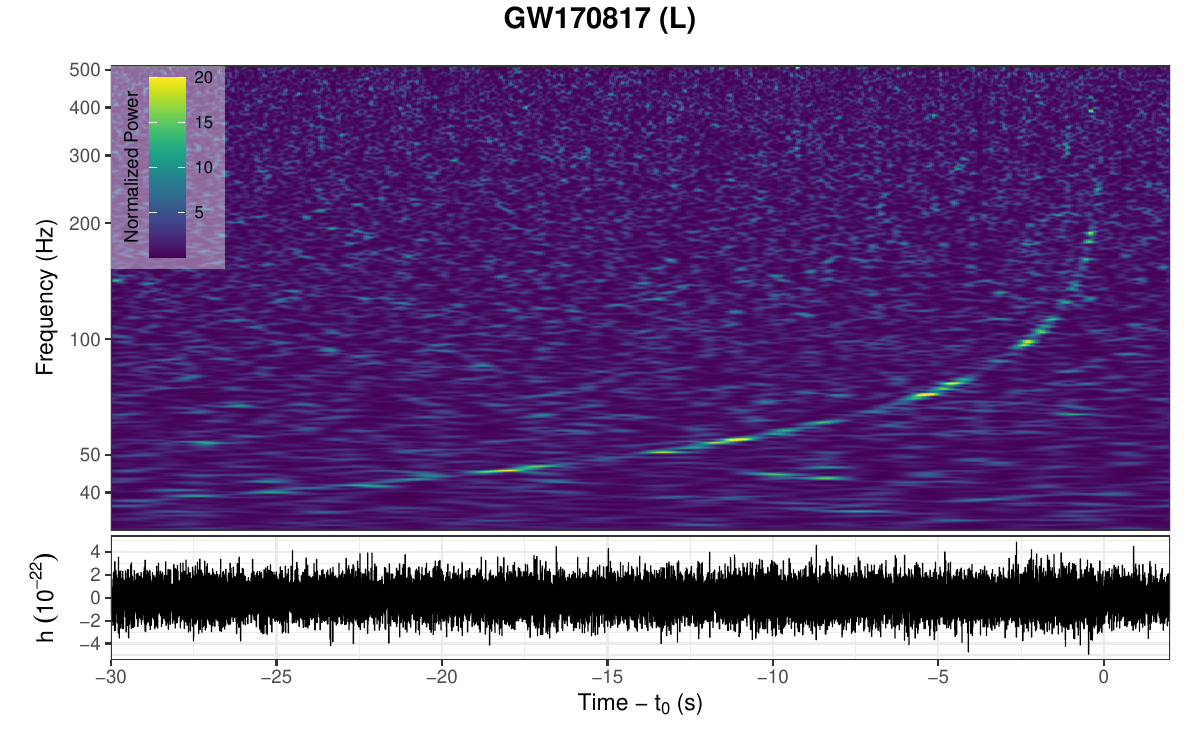}
     \end{subfigure}
     \caption{The spectrograms and oscillograms of seqARIMA-denoised LIGO data of GW150914, GW151012, and GW170817 as selected from GWTC-1 (Results for the other GWTC-1 events are shown in \autoref{fig:GWTC1rest}). The color scale for the normalized power of the spectrogram is given at the upper-left corner of each panel. The reference epochs $t_{0}$ for each case are the event time reported in GWOSC. For GW170817, the glitch-removed LIGO-L data is used.}
     \label{fig:GWTC1sel}
\end{figure*}

All 11 events in GWTC-1 can be well extracted by seqARIMA. 
In \autoref{fig:GWTC1sel}, we show the spectrograms/oscillograms of the extracted signals from three representative cases, GW150914, GW151012, and GW170817, as detected by both observatories in Hanford (H:  {\it left panels}) and Livingston (L: {\it right panels}). 
For the results of other GWTC-1 events, we have put them in the Appendix (\autoref{fig:GWTC1rest}). 

GW150914 is the first case that a GW signal was directly detected \citep{PhysRevLett.116.061102}. Its high signal-to-noise (SNR) of 26 has put it among the strongest signals of BBH merger detected so far. In Section II, we have already used this case for illustrating the feasibility of seqARIMA, in which we demonstrate that the signal of GW150914 can be clearly recovered. In the top row of \autoref{fig:GWTC1sel}, we have produced the spectrograms of this event with Q-transform for visualizing how the frequency of the signal varies over the entire process. The characteristic sweeping chirp can be clearly seen in the spectrograms. 

GW151012 is the BBH merger detected with a SNR of 10, which puts it as the weakest signal in GWTC-1 \citep{PhysRevX.9.031040}. Its low  significance as found from the initial discovery in O1 did not make it as a confirmed detection. And hence it was firstly considered as a candidate which was named as LVT151012 \citep{PhysRevX.6.041015}. With a more detailed analysis, it was found to meet the criteria of a confident detection and was subsequently re-named as GW151012. In the second row of \autoref{fig:GWTC1sel}, we show the spectrograms of the signals of GW151012 as extracted by seqARIMA. The chirp-like feature can be seen from the denoised data though it is not as clear as in the case of GW150914 because of its low significance. 


The GW signal from the event GW170817 is resulted from a merging NS-NS binary, which is the first GW event that has the counterpart detected across the whole electromagnetic spectrum \citep{PhysRevLett.119.161101,2017ApJ...848L..12A,2017ApJ...848L..13A,Goldstein_2017}. It is associated with a short $\gamma-$ray burst GRB170817A, detected by {\it Fermi} Gamma-ray Burst Monitor (GBM) 1.7 s after the coalescence \citep{Goldstein_2017}. It has provided a long-sought evidence for the link between NS-NS mergers and short $\gamma-$ray bursts. 
Unlike BBHs, the inspiral time of GW170817 is much longer. Therefore, we take this event as a test for the capability of our framework in handling a signal with a longer timescale.   
Apart from adopting a wider window in the analysis, since LIGO-L data of GW170817 suffered from the transient noise (or glitch) at the GPS time of 1187008881.389 (around 1.1~s before the coalescence), we have used the data after noise subtraction following the glitch model described in \cite{PhysRevLett.119.161101}\footnote{\href{https://www.gw-openscience.org/events/GW170817/}{https://www.gw-openscience.org/events/GW170817}}. The spectrograms of GW170817 resulted from seqARIMA denoising are shown in the bottom panels of \autoref{fig:GWTC1sel}. The inspiral and the merging process over $\sim30$~s can be clearly visualized in both data.


\subsubsection{GW190814 \& GW200105\_162426}
Apart from reproducing the GWTC-1 events, we have further tested our framework on two additional sources: GW190814 \& GW200105\_162426. These events were chosen because their inferred properties are somewhat different from those 11 events in GWTC-1. 

GW190814 was detected in the third observing run (O3) with a SNR of 25 \citep{Abbott_2020}. Parameter estimation suggests that the masses of the compact objects in their progenitor binary are highly unequal. While one component has its mass estimated as $\sim23M_{\odot}$ which is consistent with a stellar BH, the mass of the other one is likely lying in a range of $\sim2.5-3M_{\odot}$ which put it in a mass gap of being either a very massive NS or a low-mass BH. In the Fig.~1 of \citep{Abbott_2020}, we notice that the timescale of GW190814 is $\sim2-3$~s long which is different from those of GWTC-1 sources. Therefore, we have included GW190814 in our test. 

For the same reason, we have also included GW200105\_162426 (hereafter GW200105) in our experiment. It was detected by a single detector (LIGO-L) during O3 with an SNR of $\sim14$ \citep{2021ApJ...915L...5A}. It is estimated to have component masses of $\sim8.9M_{\odot}$ and $\sim1.9M_{\odot}$ which makes it likely an NS-BH binary. The signal of GW200105 shows a track of excess power with increasing frequency over $\sim3$~s in the spectrogram \citep[see Fig.~1 in][]{2021ApJ...915L...5A}. 

\begin{figure*}[htbp]
     \begin{center}
     \begin{subfigure}[b]{0.45\linewidth} 
         \includegraphics[width=\linewidth]{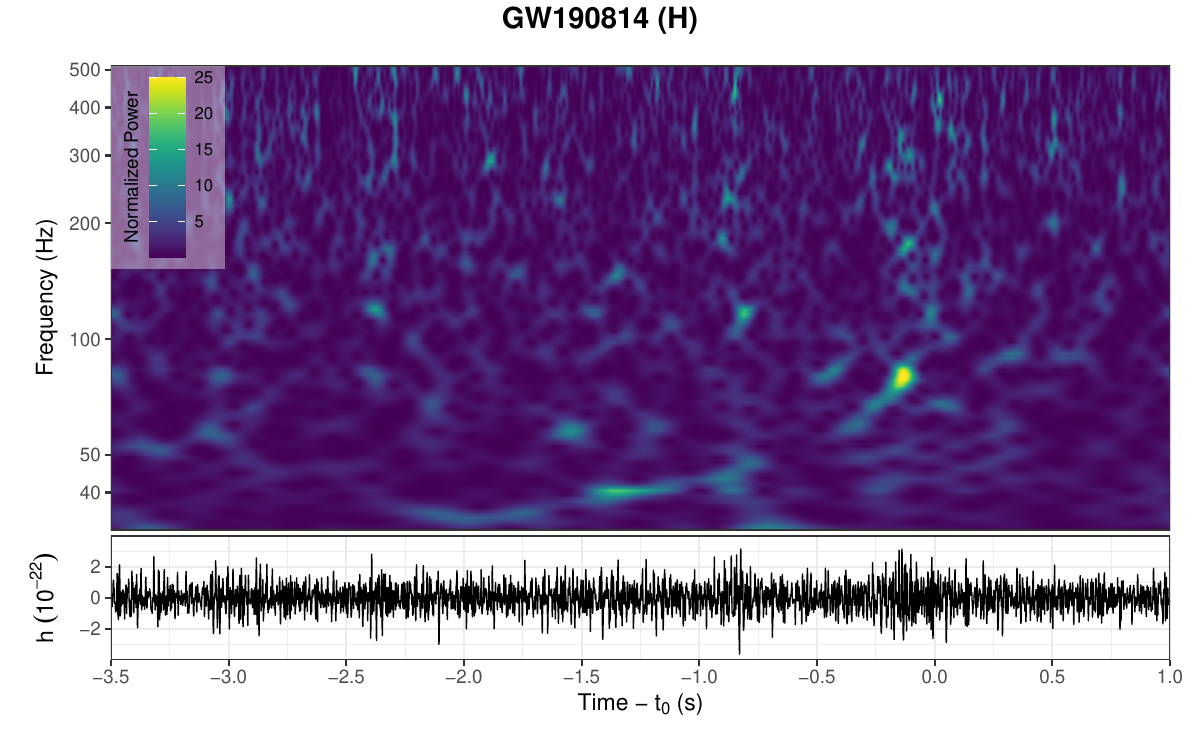}
     \end{subfigure}
     \begin{subfigure}[b]{0.45\linewidth} 
         \includegraphics[width=\linewidth]{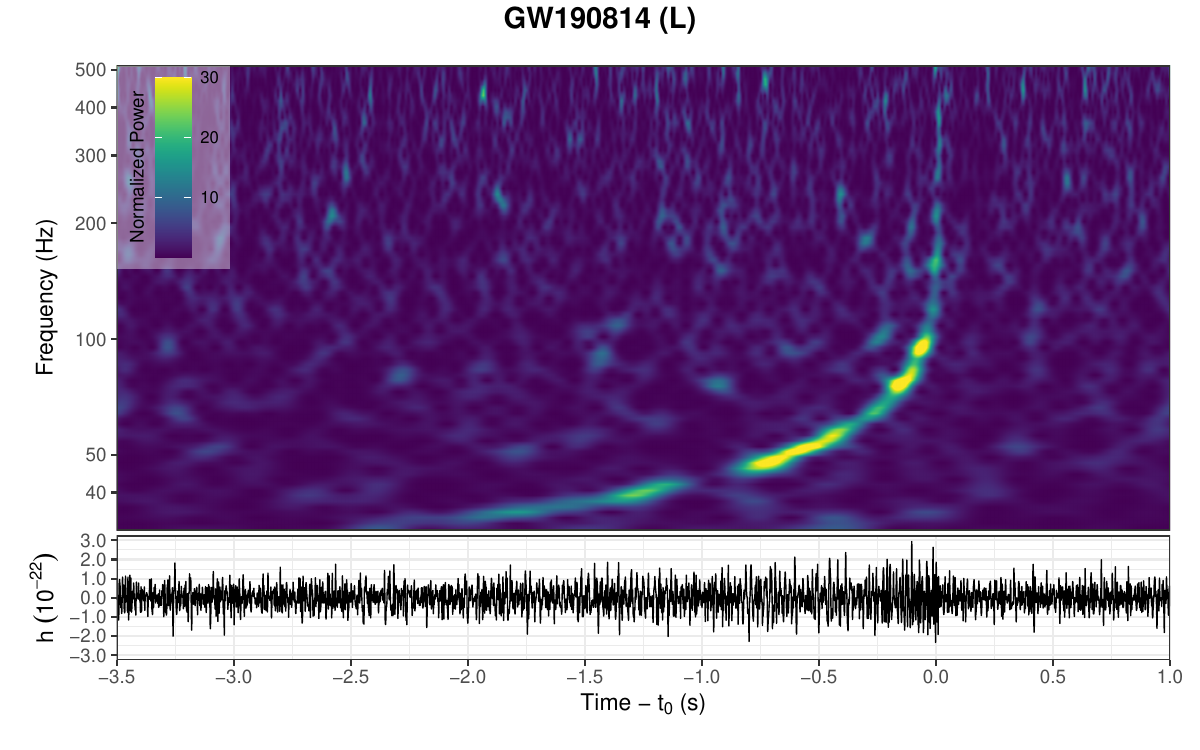}
     \end{subfigure}
     \begin{subfigure}[b]{0.45\linewidth} 
         \centering
         \includegraphics[width=\linewidth]{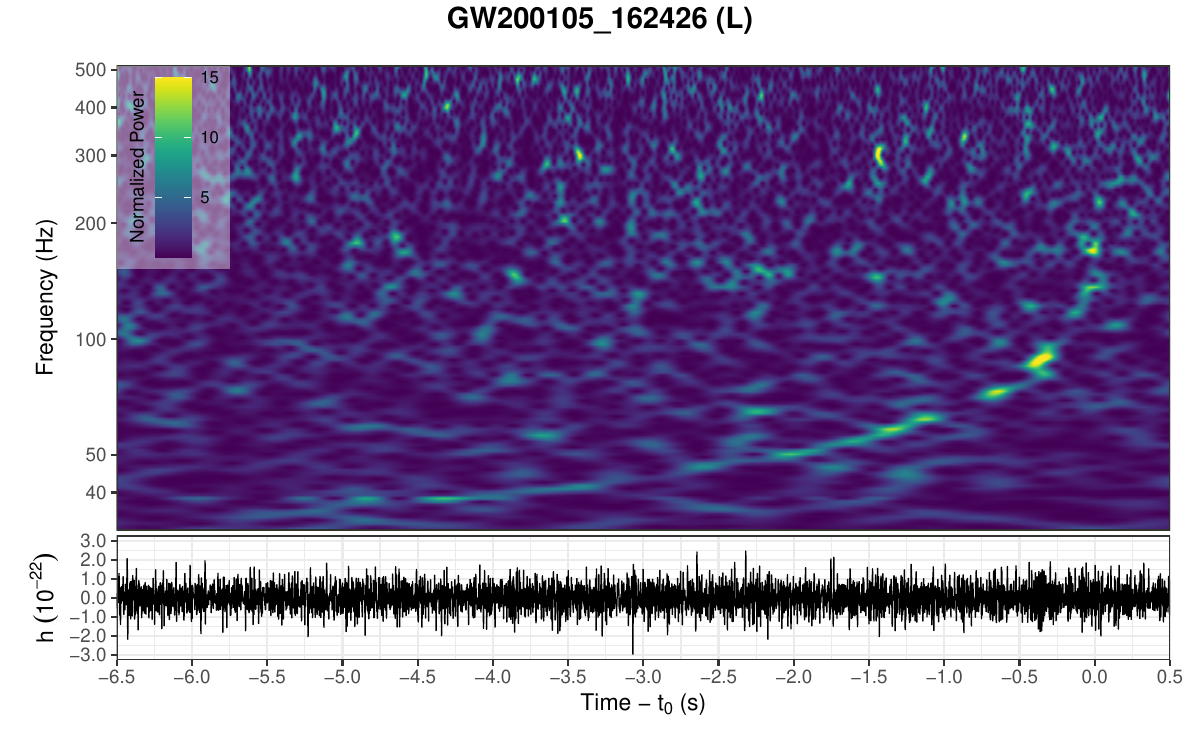}
     \end{subfigure}
     \end{center}
     \caption{The spectrograms and oscillograms of seqARIMA-denoised data of GW190814 (LIGO-H, L) and GW200105\_162426 (LIGO-L).}
     \label{fig:Extras}
\end{figure*}

In \autoref{fig:Extras}, we show the spectrograms of these two sources produced in our framework. The tracks of the signals in both cases are clearly visible. In comparing the spectrogram of GW200105 resulted from seqARIMA and the one obtained from spectral whitening as shown in Fig.~1 of \cite{2021ApJ...915L...5A}, we found that our result can attain a higher clarity which shows the inspiraling stage has a duration up to $\sim6$~s. 

In order to compare the performance of signal extraction by whitening and seqARIMA, we computed the ${\rm CCF}_{\rm max}$ and RMSE resulted from both schemes with reference to the waveforms generated by {\it pycbc} with the model of {\tt SEOBNRv4\_opt} for BBHs and {\tt IMRPhenomPv2} for GW170817 (BNS), GW190814 (Mass-gap), and GW200105 (NSBH) according to the parameters given in the corresponding literature. The results are summarized in \autoref{tab: RealData_Metrics}.
 For comparing RMSE between seqARIMA and whitening, we have seen general improvement in most cases. However, there are a few cases that the noise reduction resulted from seqARIMA are worse than that from whitening. The most notable one is from the LIGO-L data of GW150914. This might suggest that for the events with SNR sufficiently large as in the case of GW150914, seqARIMA may not have the advantage over the conventional whitening. This is also reflected by the non-monotonic behavior for the small values of $d_{L}$ in \autoref{fig:SimCCF}.
On the other hand, in terms of ${\rm CCF}_{\rm max}$ (i.e. the similarity between the extracted signals and the model), seqARIMA has shown improvement in all our tested cases. 

\section{Summary \& Future Prospects}
In this work, we have proposed a novel de-noising technique in processing GW data, which is based on autoregressive modeling. By coupling with other techniques (i.e. integrated process, EoA), we have developed a framework we refer as seqARIMA pipeline (cf. \autoref{fig:FlowChart}). The effects of each component in the pipeline have been investigated (see \autoref{fig:ARIMAsteps} and \autoref{sec:integProc}-\autoref{sec:bpProc} for details). We have tested the performance of our proposed framework with a series of experiments. 

We have examined the ability of seqARIMA pipeline in extracting the simulated GW signal with varying waveform and distance. By comparing the noise-subtracted time series and the injected signal (\autoref{fig:SimBetas}), we have computed CCF$_{\rm max}$ and RMSE resulted from both seqARIMA and whitening process. At larger distance, we found that seqARIMA can attain a higher CCF$_{\rm max}$ and lower RMSE than those resulted from whitening (\autoref{fig:SimCCF}).

We have also applied our method in extracting a number of known GW events from the LIGO data. All 11 events cataloged in GWTC-1 can be well recovered by seqARIMA (\autoref{fig:GWTC1sel} \& \autoref{fig:GWTC1rest}). We have further tested the method in two additional sources GW190814 (mass-gap object) and GW200105 (NS-BH merger), which have the timescale of their GW signals different from those in GWTC-1. We showed their signals can also be successfully extracted (\autoref{fig:Extras}). 

We have further compared the CCF$_{\rm max}$ and RMSE resulted from both seqARIMA and whitening by comparing the noise-subtracted time series of these events with the model waveforms generated in accordance with the parameters specified in the corresponding literature (see \autoref{tab: RealData_Metrics}). We found that seqARIMA generally yields improvement over whitening in terms of these performance metrics. 

We have demonstrated that seqARIMA can enhance the noise suppression and therefore it is capable to provide an alternative to the conventional frequency-domain whitening process. For further improving the denoising performance, seqARIMA can be coupled with deep learning. Many recent studies have investigated the feasibility of denoising the GW data with deep neural network and showed that this can significantly suppress the noise and recover the signal \citep[e.g.][]{8683061,WEI2020135081,ren2022intelligent,2023CmPhy...6..212Z}. We notice that these recent studies remain using whitening as a preprocessing procedure. Therefore, it will be encouraging to explore whether combining seqARIMA with these machine-learning based architectures can boost the denoising performance to a further extent. By substituting whitening with our proposed method, dedicated studies can also explore whether parameter estimation can also be benefited from seqARIMA.

 We can also consider the feasibility of incorporating seqARIMA into a template-free low-latency detection pipeline. Since whitening can be a dominant source for the latency, it is desirable to reduce the computational cost in this stage \citep[e.g.][]{2018PhRvD..97j3009T}. However, the conventional frequency-domain whitening process do not have many degree of freedom for improving the computational efficiency. On the other hand, the complexity of seqARIMA can be controlled by the hyper-parameters $p_{\rm max}$ and $q_{\rm max}$, which gives the flexibility of this process. For example, in trading off the fidelity of the extracted signal, a low $p_{\rm max}$ can result in a more efficient modeling.  Therefore, one can examine whether seqARIMA can be adopted in a candidate identification pipeline. With the improved noise subtraction, the signal from a CBC or burst event can possibly be identified as a cluster of bright pixels in the spectrograms \citep[e.g.][]{Honda_2008,PhysRevD.88.083010} which allows a GW event candidate to be detected without a priori knowledge of its waveform. A quantitative analysis on the execution speed of our proposed framework will be important for examining the capability of rapid real-time processing.

\begin{acknowledgments}
The authors would like to thank Dr. Wang He for his valuable comments for improving the quality of this work. S.K. is supported by the National Research Foundation of Korea grant 2022R1F1A1073952. C.Y.H. is supported by the research fund of Chungnam National University and by the National Research Foundation of Korea grant 2022R1F1A1073952. A.K.H.K. is supported by the National Science and Technology Council of Taiwan through grants 111-2112-M-007-020 and 112-2112-M-007-042. L.C.C.L is supported by NSTC of Taiwan through grant Nos. 110-2112-M-006-006-MY3 and 112-2811-M-006-019. K.L.L. is supported by the National Science and Technology Council of the Republic of China (Taiwan) through grant 111-2636-M-006-024, and he is also a Yushan Young Fellow supported by the Ministry of Education of the Republic of China (Taiwan). J.Y. and A.P.L. is supported by the Science and Technology Development Fund, Macau SAR (No. 0079/2019/A2).
\end{acknowledgments}



\bibliography{bibtex} 


\appendix

\section{Spectrograms and oscillograms of GWTC-1 events produced by our framework}
\begin{figure*}[htbp]
     \centering
     \begin{subfigure}[b]{0.45\linewidth}
         \centering
         \includegraphics[width=\linewidth]{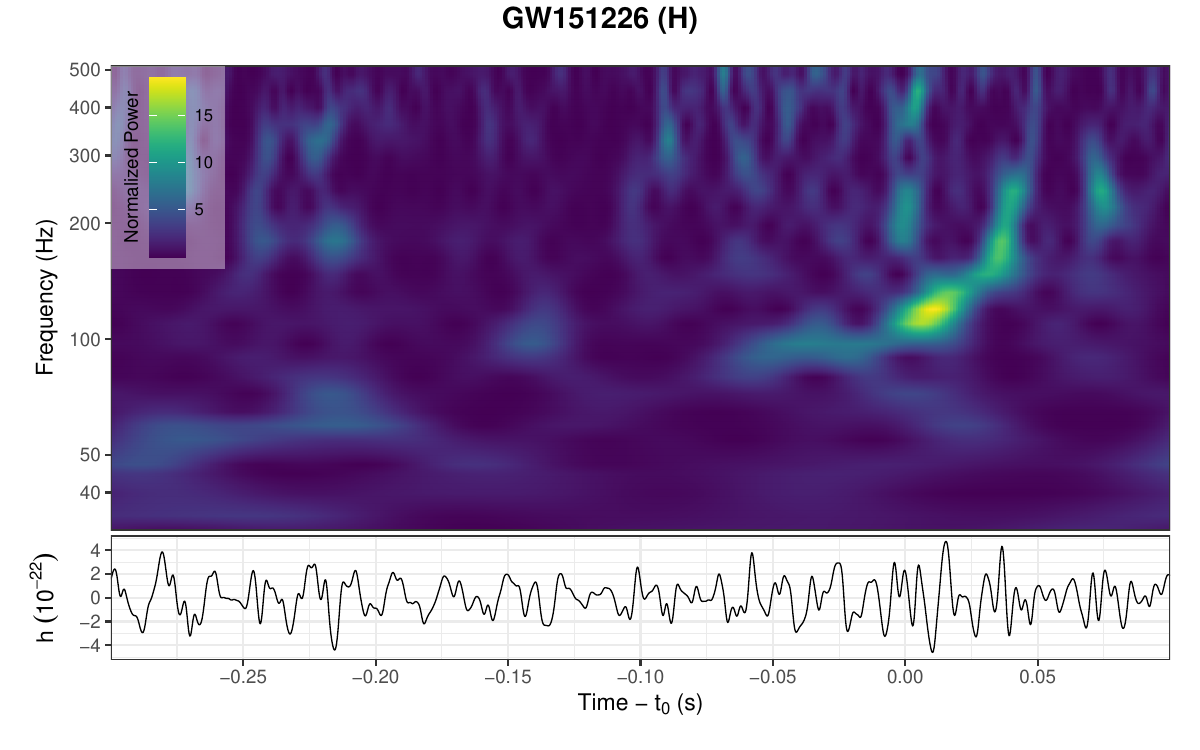}
     \end{subfigure}
     \begin{subfigure}[b]{0.45\linewidth}
         \centering
         \includegraphics[width=\linewidth]{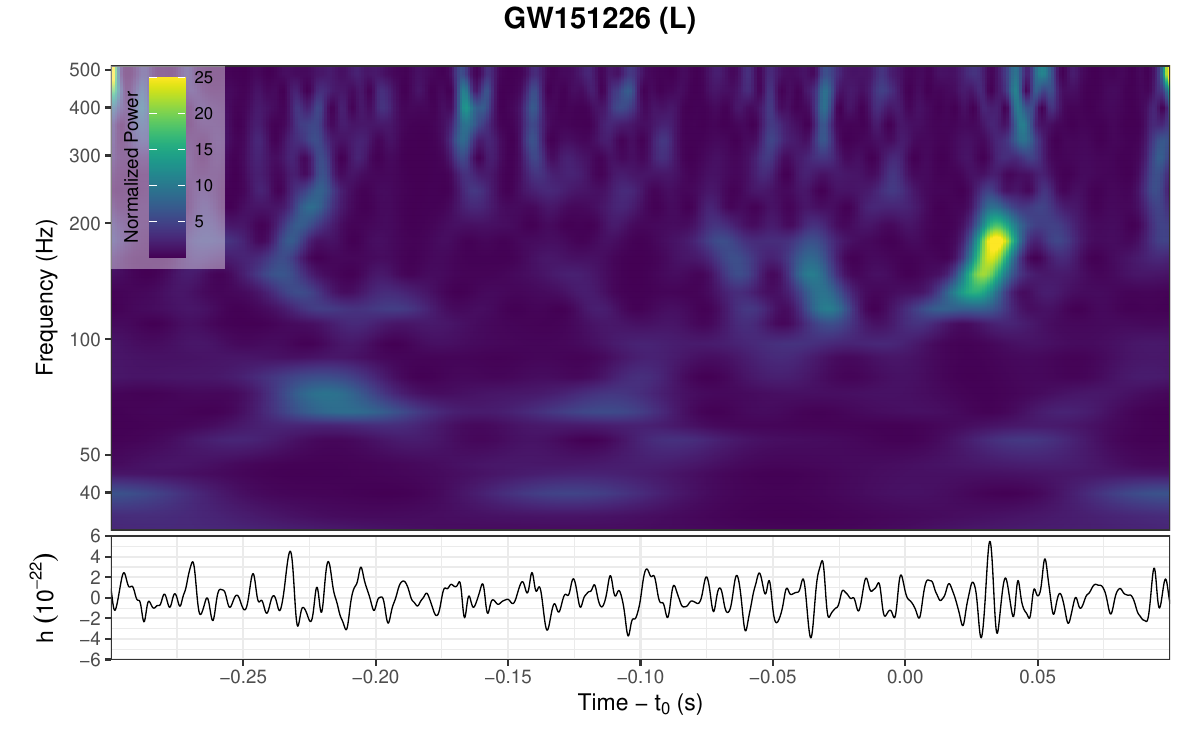}
     \end{subfigure}

     \begin{subfigure}[b]{0.45\linewidth}
         \centering
         \includegraphics[width=\linewidth]{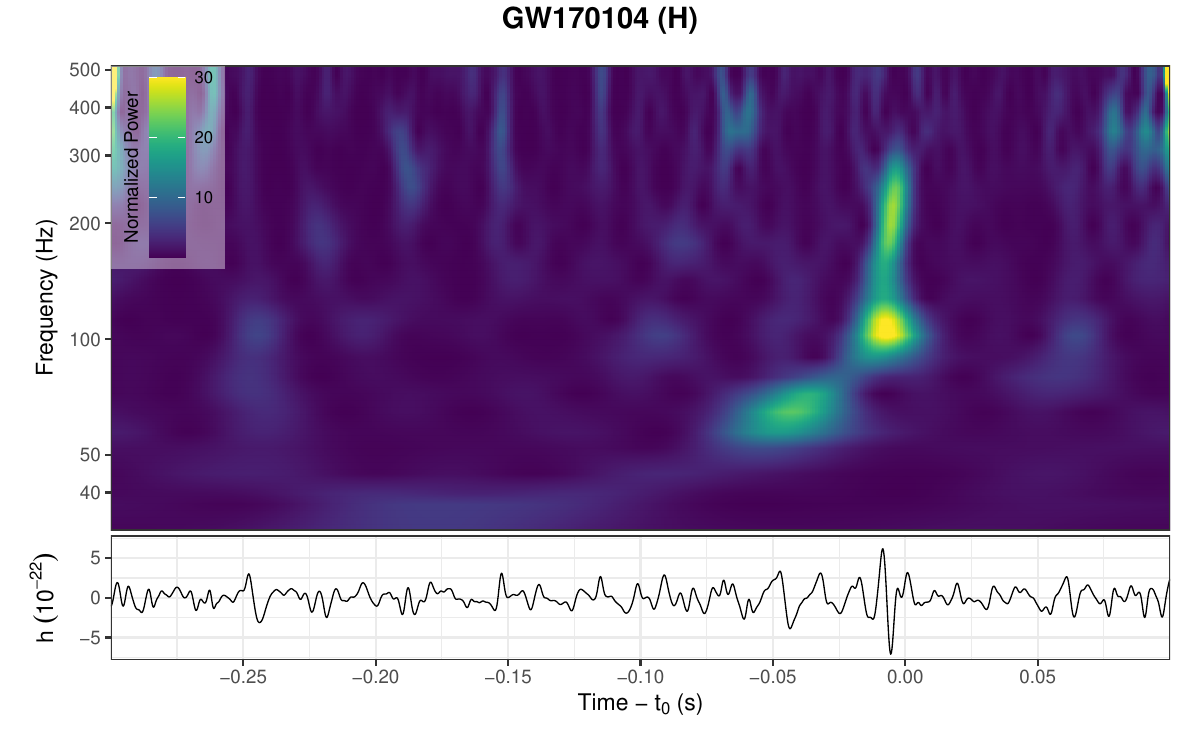}
     \end{subfigure}
     \begin{subfigure}[b]{0.45\linewidth}
         \centering
         \includegraphics[width=\linewidth]{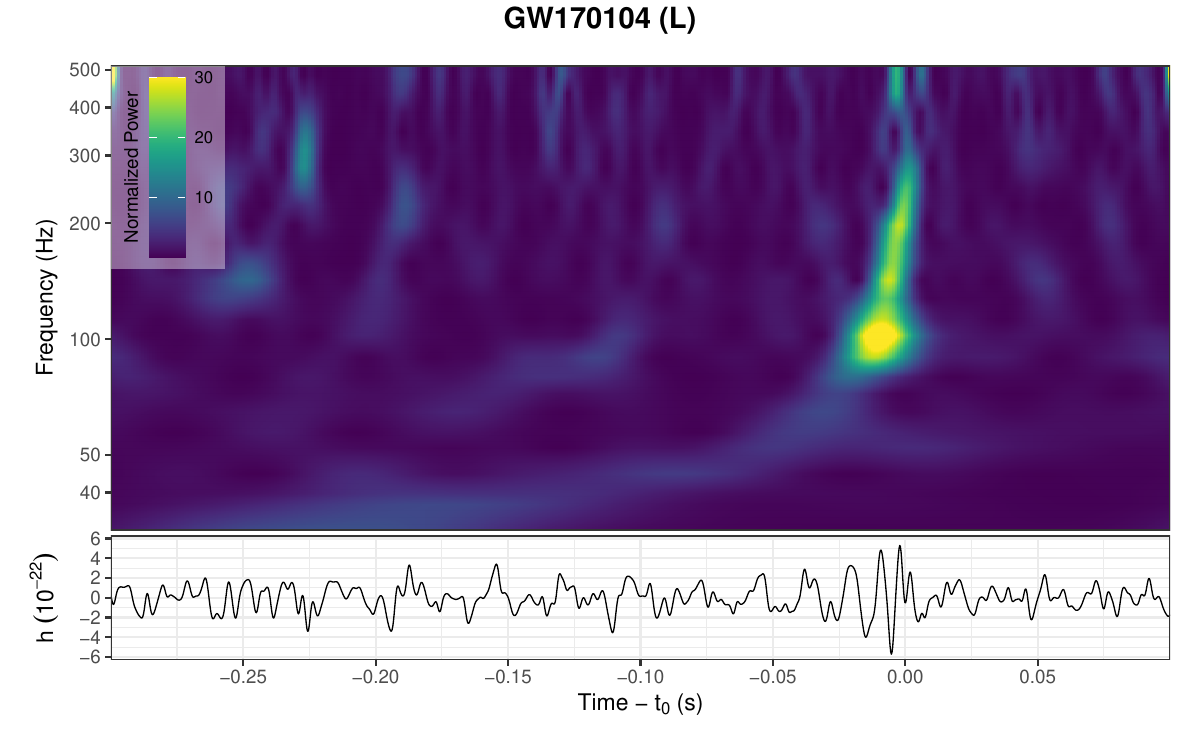}
     \end{subfigure}
     
     \begin{subfigure}[b]{0.45\linewidth}
         \centering
         \includegraphics[width=\linewidth]{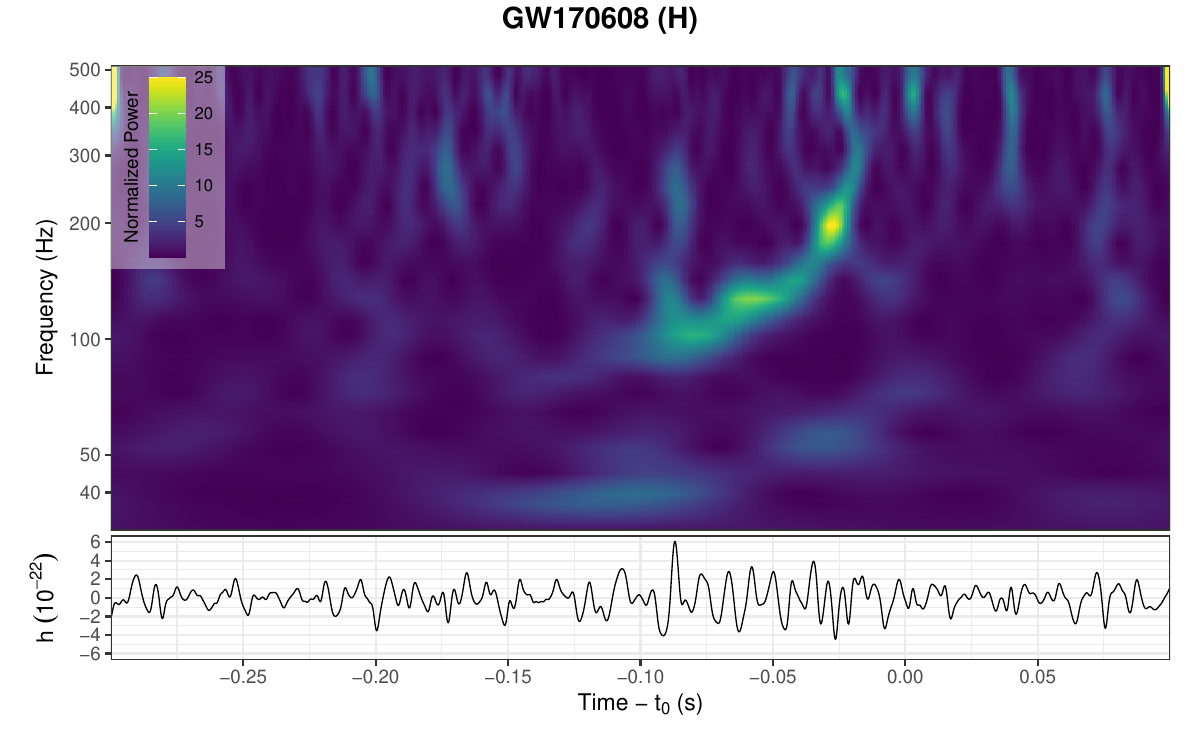}
     \end{subfigure}
     \begin{subfigure}[b]{0.45\linewidth}
         \centering
         \includegraphics[width=\linewidth]{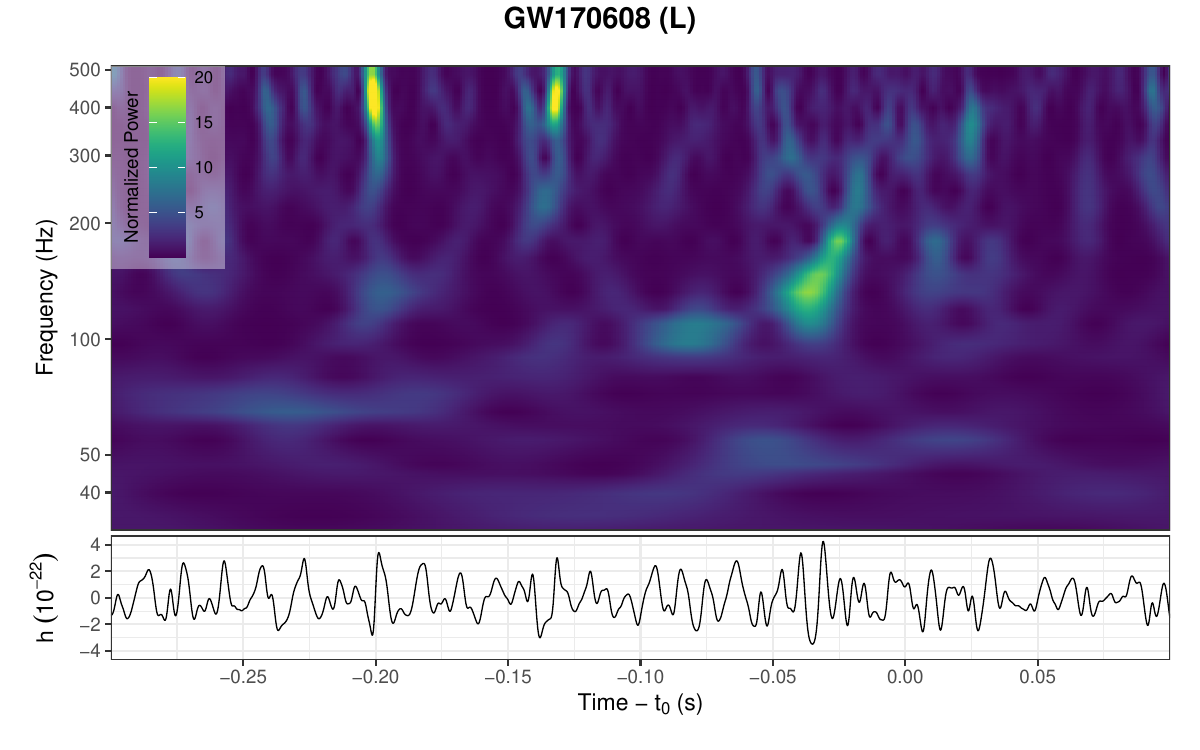}
     \end{subfigure}

     \begin{subfigure}[b]{0.45\linewidth}
         \centering
         \includegraphics[width=\linewidth]{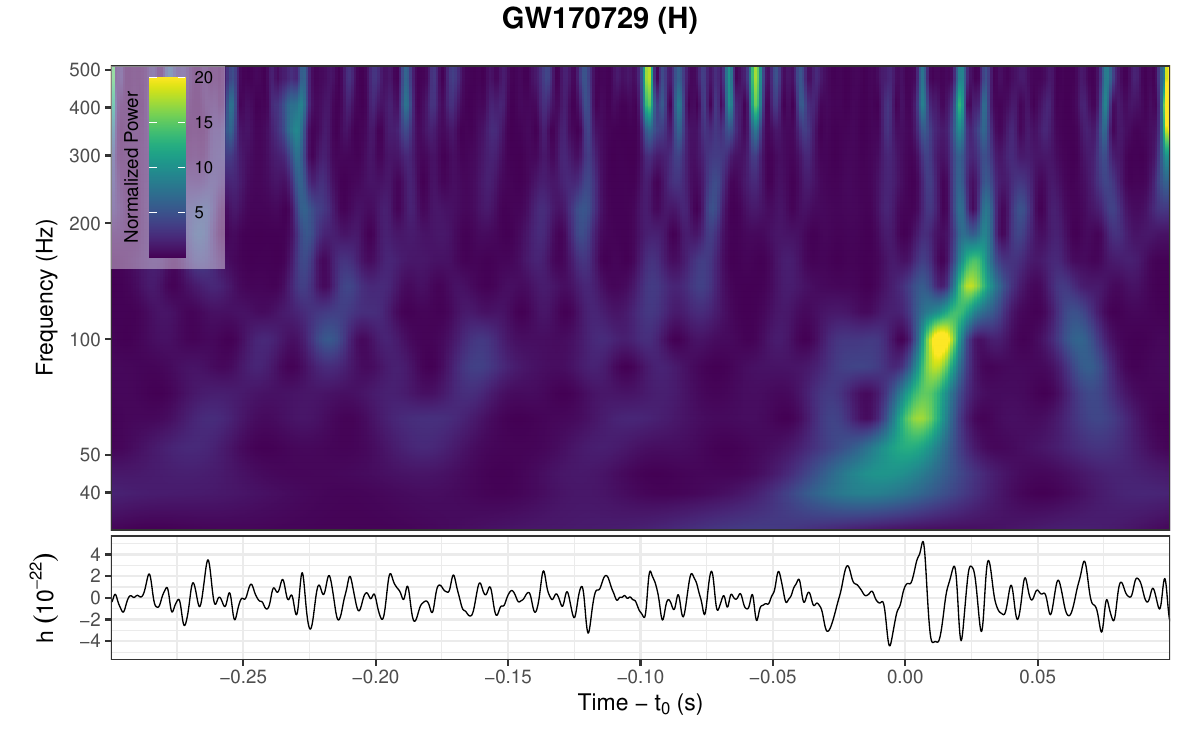}
     \end{subfigure}
     \begin{subfigure}[b]{0.45\linewidth}
         \centering
         \includegraphics[width=\linewidth]{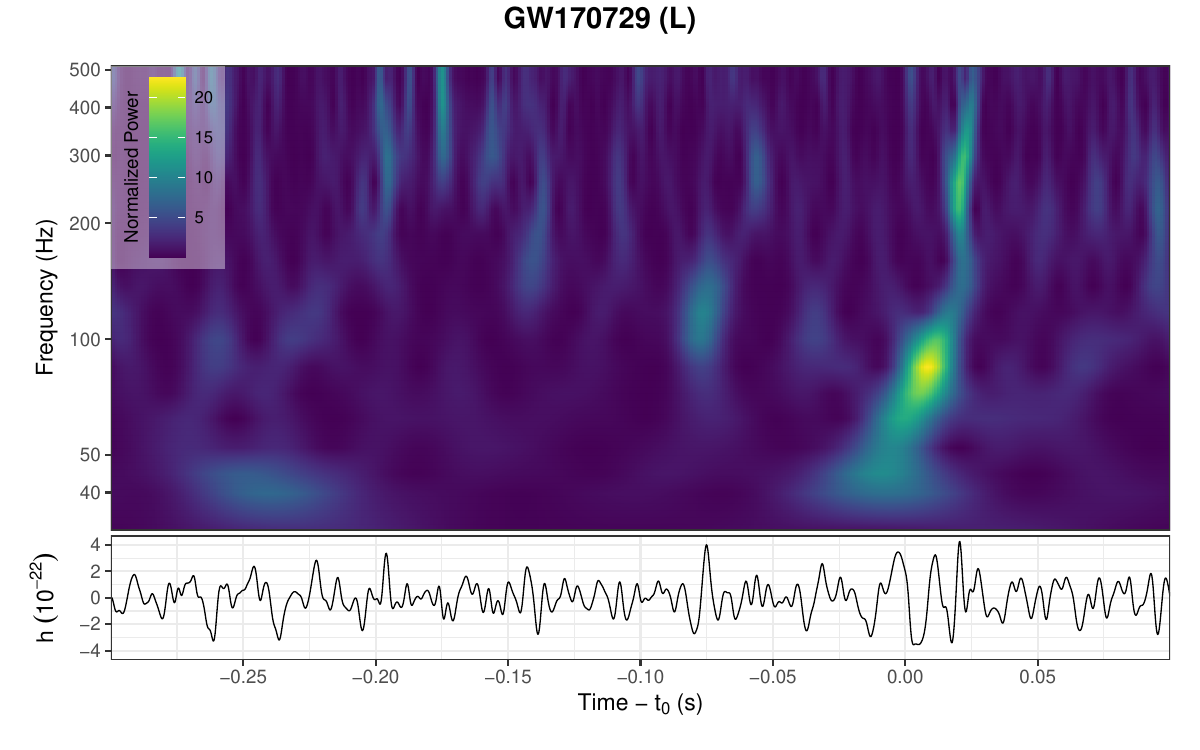}
     \end{subfigure}
     \caption{The spectrograms and oscillograms of seqARIMA-denoised LIGO data of GWTC-1 sources which are not shown in \autoref{fig:GWTC1sel}. }
     \label{fig:GWTC1rest}
\end{figure*}%
\begin{figure*}[htbp]\ContinuedFloat
\centering
     \begin{subfigure}[b]{0.45\linewidth}
         \centering
         \includegraphics[width=\linewidth]{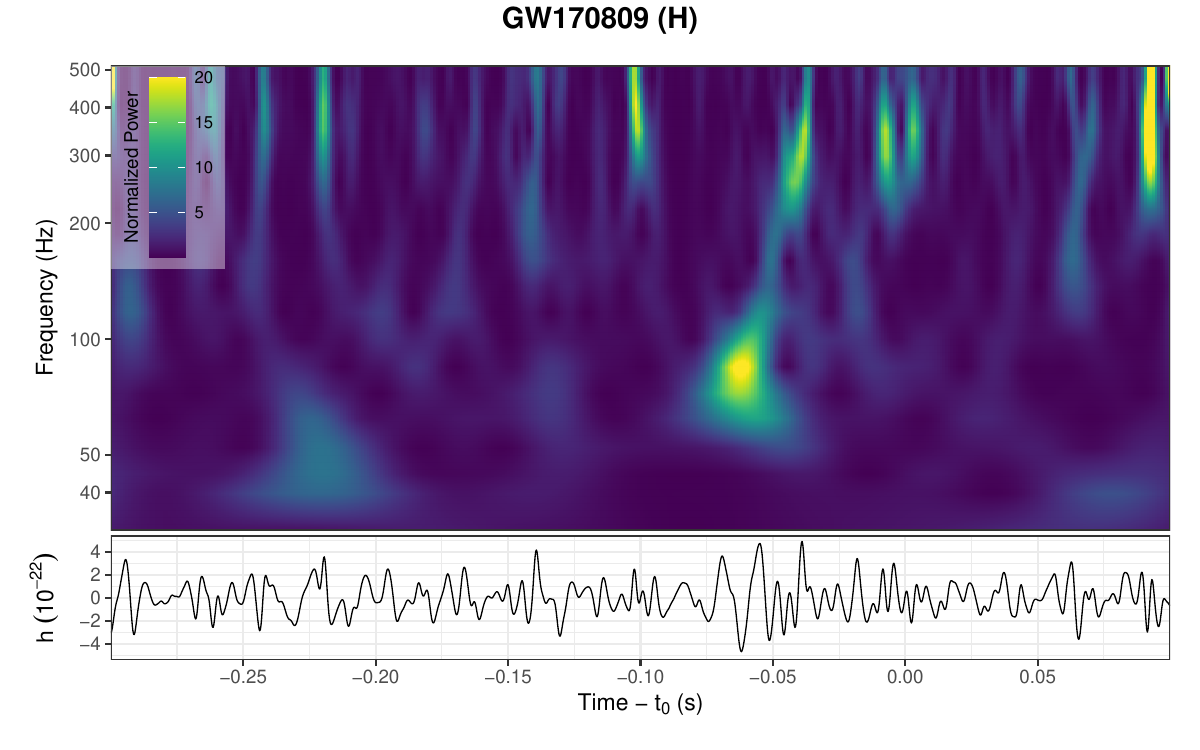}
     \end{subfigure}
     \begin{subfigure}[b]{0.45\linewidth}
         \centering
         \includegraphics[width=\linewidth]{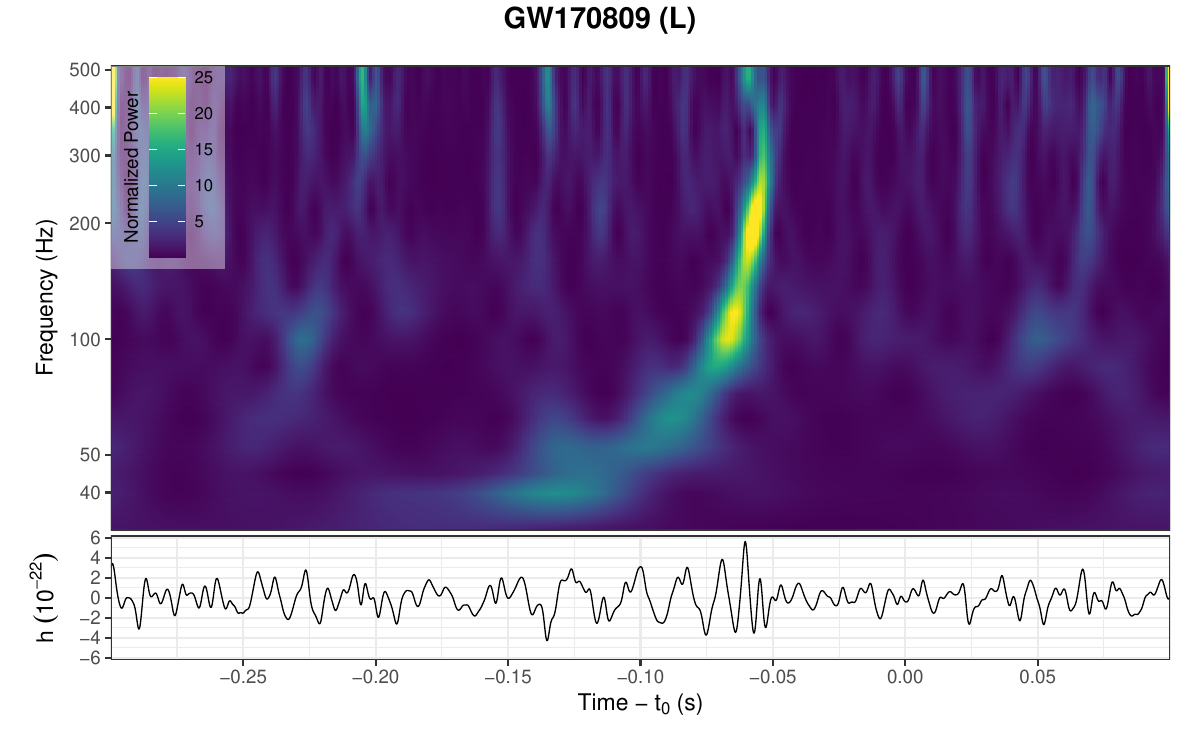}
     \end{subfigure}

     \begin{subfigure}[b]{0.45\linewidth}
         \centering
         \includegraphics[width=\linewidth]{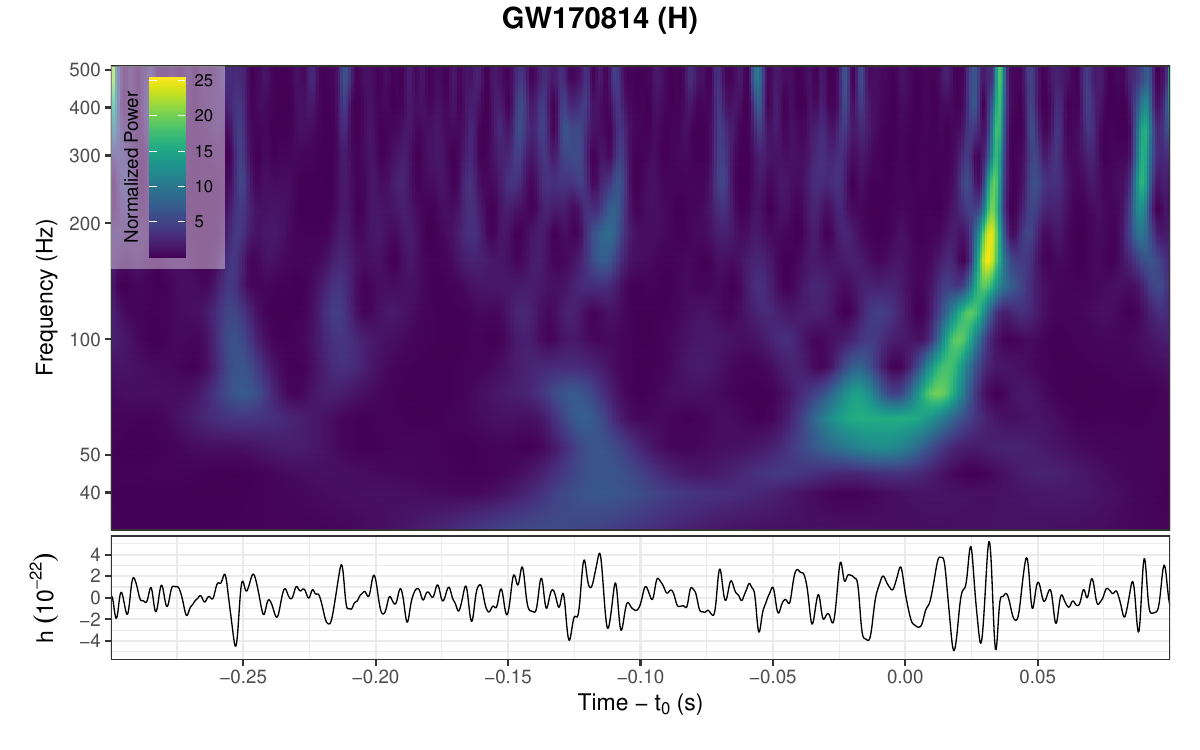}
     \end{subfigure}
     \begin{subfigure}[b]{0.45\linewidth}
         \centering
         \includegraphics[width=\linewidth]{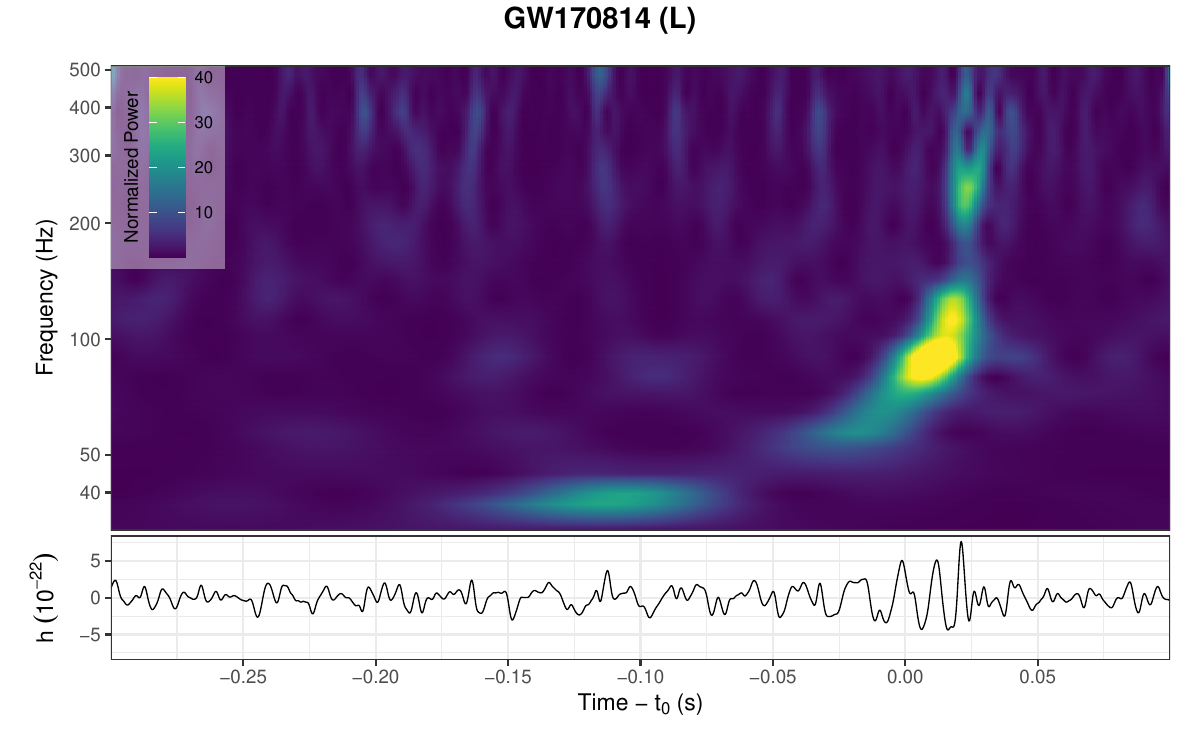}
     \end{subfigure}

     \begin{subfigure}[b]{0.45\linewidth}
         \centering
         \includegraphics[width=\linewidth]{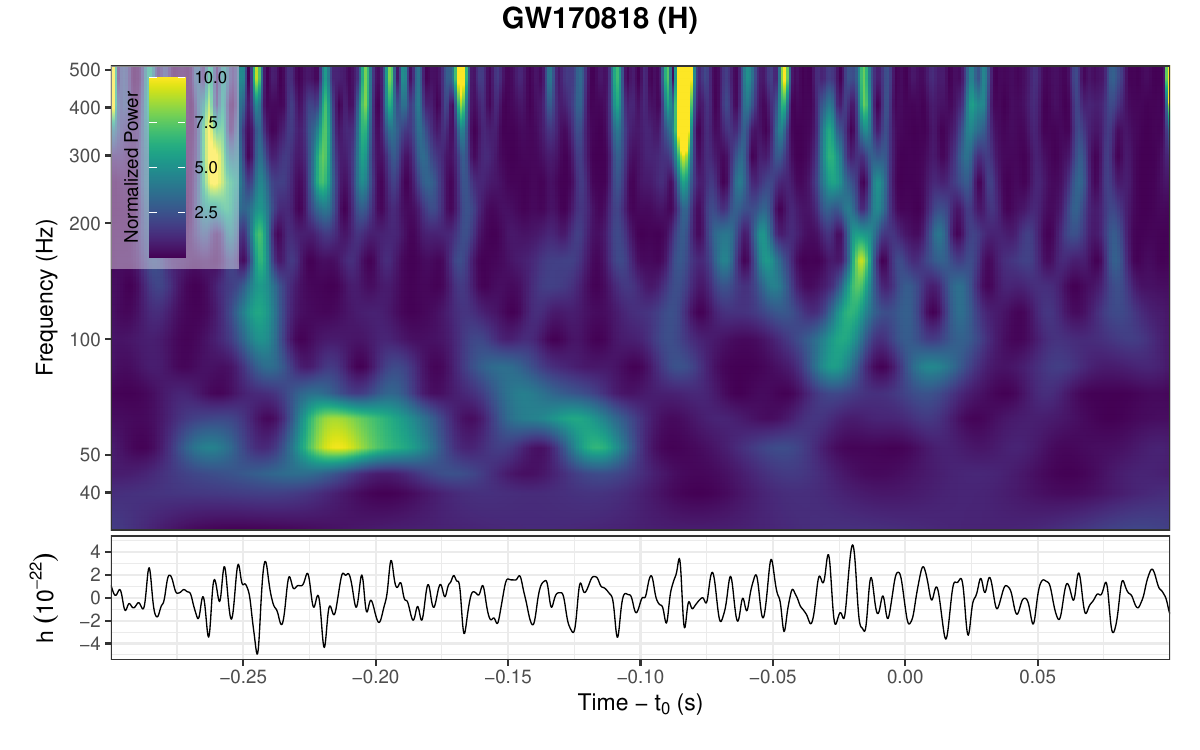}
     \end{subfigure}
     \begin{subfigure}[b]{0.45\linewidth}
         \centering
         \includegraphics[width=\linewidth]{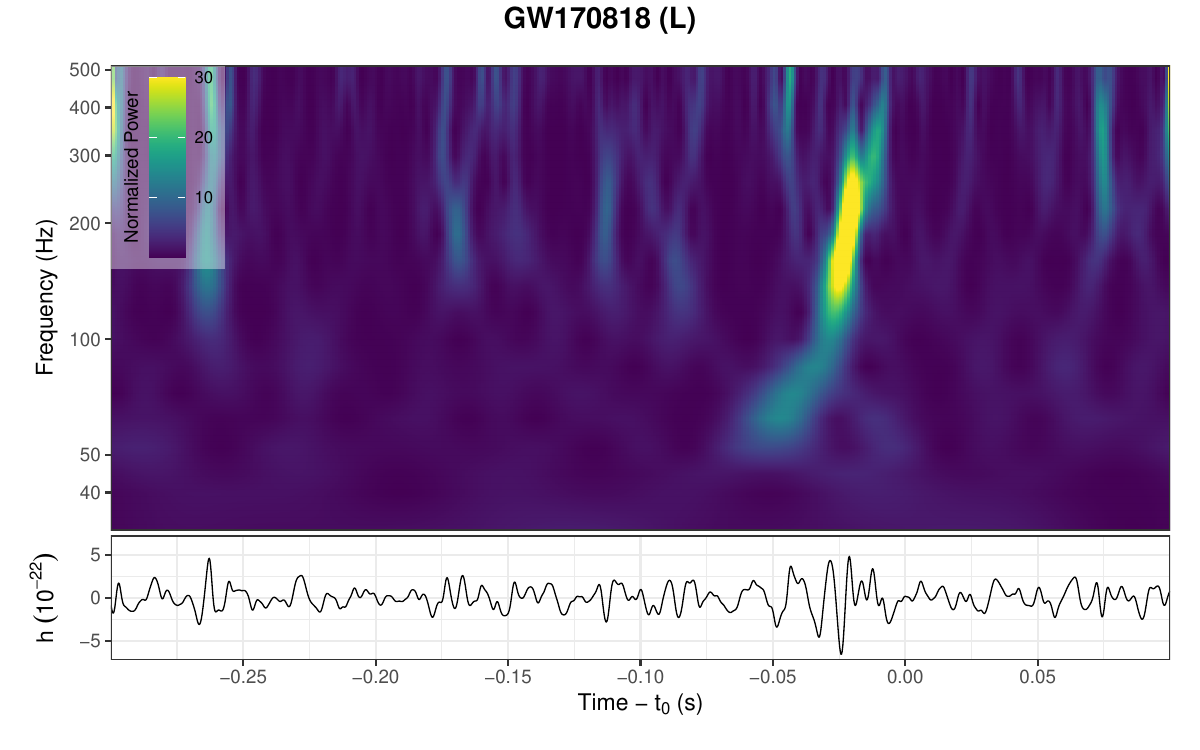}
     \end{subfigure}

     \begin{subfigure}[b]{0.45\linewidth}
         \centering
         \includegraphics[width=\linewidth]{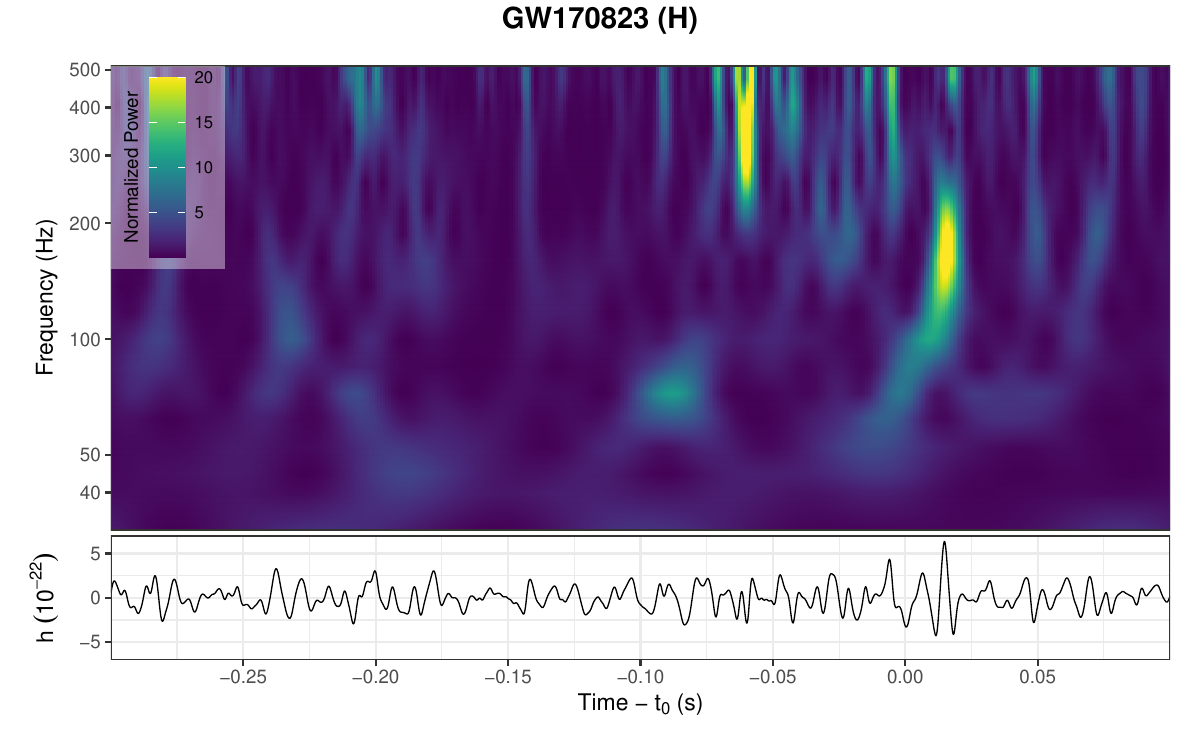}
     \end{subfigure}
     \begin{subfigure}[b]{0.45\linewidth}
         \centering
         \includegraphics[width=\linewidth]{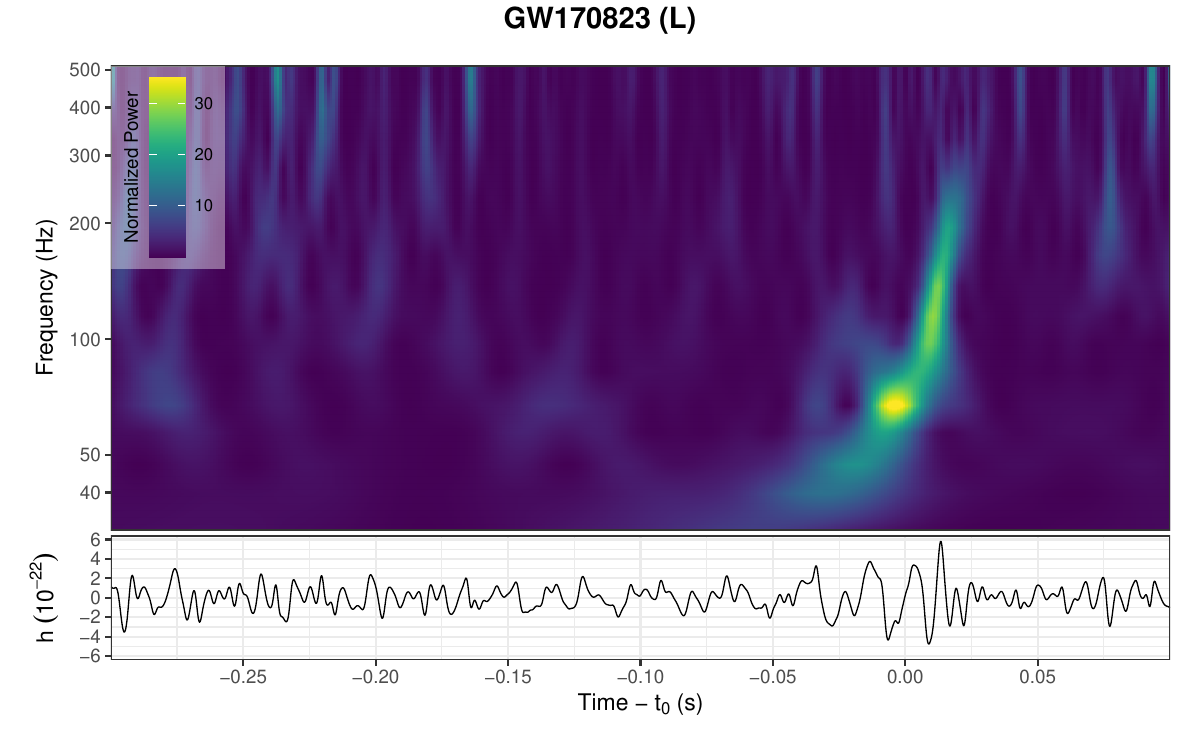}
     \end{subfigure}
     \caption[]{The spectrograms and oscillograms of seqARIMA-denoised LIGO data of GWTC-1 sources which are not shown in \autoref{fig:GWTC1sel}.}
\end{figure*}


\end{document}